\documentclass[apj]{emulateapj}

\usepackage{amsmath}	% Advanced maths commands
\usepackage{amssymb}	% Extra maths symbols
\usepackage{multirow}
\usepackage{subfigure}
\usepackage{booktabs}
\usepackage{color}
\def\wise{\textit{WISE}}
\def\madcows{MaDCoWS}
\def\spz{\textit{Spitzer}}
\def\as{$^{\prime\prime}$}
\shorttitle{Radio-AGN in Massive Galaxy Clusters}
\shortauthors{E. Moravec et al.}

\begin{document}

\title{The Massive and Distant Clusters of WISE Survey VII: The Environments and Properties of Radio Galaxies in Clusters at $z\sim$1}

\author{Emily Moravec\altaffilmark{1},
Anthony H. Gonzalez\altaffilmark{1}, 
Daniel Stern\altaffilmark{2},
Tracy Clarke\altaffilmark{3},
Mark Brodwin\altaffilmark{4},
Bandon Decker\altaffilmark{4},
Peter R. M. Eisenhardt\altaffilmark{2},
Wenli Mo\altaffilmark{1}, 
Alexandra Pope\altaffilmark{5},
Spencer A. Stanford\altaffilmark{6}, 
and
Dominika Wylezalek\altaffilmark{7}
}

\affil{$^1$Department of Astronomy, University of Florida, Gainesville, FL, 32611, USA}
\affil{$^2$Jet Propulsion Laboratory, California Institute of Technology, Pasadena, CA 91109, USA}
\affil{$^3$Naval  Research  Laboratory,  Code  7213,  Washington,  DC,
20375, USA}
\affil{$^4$Department of Physics and Astronomy, University of Missouri, 5110 Rockhill Road, Kansas City, MO 64110, USA}
\affil{$^5$Department of Astronomy, University of Massachusetts, 710 North Pleasant Street Amherst, MA 01003, USA}
\affil{$^6$Department of Physics, University of California, Davis, One Shields Avenue, Davis, CA 95616, USA}
\affil{$^7$European Southern Observatory, Karl-Schwarzschild-Str. 2, 85748 Garching bei M\"unchen, Germany}

%%%%%%%%%%%%%%%%%%%%%%%%%%%%%%%%%%%%%
% Abstract of the paper
\begin{abstract}
We present the results from a study with NSF's Karl G. Jansky Very Large Array (VLA) to determine the radio morphologies of extended radio sources and the properties of their host galaxies in 50 massive galaxy clusters at $z\sim1$. We find a majority of the radio morphologies to be Fanaroff-Riley (FR) type IIs. By analyzing the infrared counterparts of the radio sources, we find that $\sim$40\% of the host galaxies are the candidate brightest cluster galaxy (BCG) and $\sim$83\% are consistent with being one of the top six most massive galaxies in the cluster. We investigate the role of environmental factors on the radio-loud AGN population by examining correlations between environmental and radio-galaxy properties. We find that the highest stellar mass hosts ($M_{*} \gtrsim$ 4$\times 10^{11} M_{\odot}$) are confined to the cluster center and host compact jets. There is evidence for an increase in the size of the jets with cluster-centric radius, which may be attributed to the decreased ICM pressure confinement with increasing radius. Besides this correlation, there are no other significant correlations between the properties of the radio-AGN (luminosity, morphology, or size) and environmental properties (cluster richness and location within the cluster). The fact that there are more AGN in the cluster environment than the field at this epoch, combined with the lack of strong correlation between galaxy and environmental properties, argues that the cluster environment fosters radio activity but does not solely drive the evolution of these sources at this redshift.
\end{abstract}

\maketitle

%%%%%%%%%%%%%%%%%%%%%%%%%%%%%%%%%%%%%%%%%%%%%%%%%%
%%%%%%%%%%%%%%%%% BODY OF PAPER %%%%%%%%%%%%%%%%%%
\section{Introduction}\label{intro}
It has long been understood that environment is an important component of galaxy evolution. In particular, the co-evolution of and connection between active galactic nuclei (AGN) and dense environments has been an active area of investigation in recent years. On one hand, the intracluster medium (ICM) in galaxy clusters can affect the AGN within the cluster in various ways. For example, the hot ICM in dense environments can impact AGN activity as the gas can either become fuel for the AGN as it cools and flows toward the central galaxies, or if prevented from cooling, it will suppress AGN fueling and star formation in the central galaxy \citep{McNamara07,ODea08,Blanton10,Prasad15,McDonald18}. Additionally, the ICM can confine AGN outflows, which is showcased by the many examples of coincident radio jets and X-ray cavities as the AGN jets inject the ICM that confines them (e.g. Hydra A: \citealt{McNamara00}, Perseus: \citealt{Fabian00}, Abell 2199: \citealt{Johnstone02}, Centaurus: \citealt{Fabian05}, M87/Virgo: \citealt{Forman07}, etc.). As a last example, the relative motion of the AGN host galaxy with respect to the ICM can distort AGN outflows (e.g. bent-tail radio sources in galaxy clusters as seen in \citealt{PM17}). On the other hand, feedback from AGN jets can have a significant mechanical and thermal impact upon the surrounding gas, preventing it from cooling \citep{McNamara07,Gaspari11,McDonald13,HL15,Li17,Yang19}. Through bursts of AGN activity the jets inject energy into surrounding gas through shock waves, dissipation of sound waves, and buoyantly rising bubbles that have been inflated by the jets (see \citealt{Blanton10,Fabian12} for a review).

Two major modes of AGN feedback have been identified and differentiated by the energy outflow near the black hole: `quasar-mode' and `radio-mode' (see \citealt{Fabian12} for a review).~During quasar-mode feedback (also known as radiative or wind mode) the AGN is radiatively efficient and its outflows can expel interstellar gas from the host galaxy, slowing the infall of matter into both the galaxy and its central supermassive black hole \citep[SMBH,][]{Kauffmann00, Granato04, DiMatteo05, Hopkins10}. On the other hand, during radio-mode feedback (also known as kinetic, radio-jet, or maintenance mode) the AGN is radiatively inefficient and is capable of driving powerful, kpc-scale jets. Despite this prevailing nomenclature, there also exists a substantial population of radio-loud quasars, accounting for 10-15\% of the luminous quasar population, where the central engine is both radiatively efficient and driving a powerful, kpc-scale jet \citep[e.g.,][]{Stern00}. 

In this work, we focus on radio galaxies which are AGN that are emitting synchrotron radiation observable in the radio wavelengths (radio-AGN). The link between radio galaxies and dense environments is now well-established. The association of radio galaxies with galaxy clusters dates back to the 1950s \citep{Minkowski60}. More recently, studies have used radio AGN to successfully identify rich, high-redshift ($z\gtrsim1$) clusters, cluster candidates, and protoclusters: the Clusters Around Radio-Loud AGN program \citep[CARLA,][]{Wylezalek13,Wylezalek14,Noirot16,Noirot18}, the Clusters Occupied by Bent Radio AGN \citep[COBRA,][]{PM17}, \cite{Castignani14}, and \cite{Rigby14}. 

Beyond the obvious link between the presence of a radio-AGN and a dense environment, studying correlations between the characteristics of the radio-AGN and their host environments can give further insight into the complexities of the co-evolution of radio-AGN and their environments. A wide range of correlations between the characteristics of the environment and radio-AGN have been found. Some of these correlations include: the number of radio-selected AGN is 2.5 times higher in clusters than in the field at $z\sim$1 \citep[]{Mo18}, the mean power and mean physical size of the brightest associated AGN increases with cluster richness \citep[]{Ineson15,Croston19}, the lowest luminosity sources are likely to be at a large distance from the group/cluster center, while the most radio-luminous AGN are typically close to the center \citep[]{Croston19}, the radio activity of the brightest cluster galaxy (BCG) depends on redshift \citep[]{Gralla11} and host-galaxy mass \citep[]{Best07}, and that the morphology of bent tail radio sources depends on distance from the cluster center \citep[]{Silverstein18,Garon19}. These correlations and others indicate that the evolution of radio-AGN and their environment are indeed intertwined, but further investigations are needed at high-redshift in order to determine the evolutionary progression.

Recent studies have shown that the number of AGN in clusters and the ratio of AGN to cluster galaxies increase with increasing redshift \citep[]{Galametz09,Martini13,Alberts16,Bufanda17,Mo18} which further showcases a connection between environment and AGN. The epoch of $z\sim$1 is an era when clusters are still assembling and evolving. It is thus an ideal epoch to directly observe the role of radio-mode feedback on cluster evolution. The Massive and Distant Cluster of \wise\ Survey (MaDCoWS) is the largest galaxy cluster sample at this redshift range, with a catalog of $\sim$2300 galaxy clusters at $z\sim$1 \citep[]{Gonzalez19}. MaDCoWS provides an ideal sample with which to investigate the connection between AGN and dense environments at high redshift.

Using MaDCoWS, there have been two works that have investigated the distribution of AGN and properties of the AGN population associated with $z\sim$1 galaxy clusters. \cite{Mo18} studied the cluster AGN population in MaDCoWS galaxy clusters as a function of central cluster distance and cluster richness. \cite{Mo18} find a distinct overdensity of AGNs within 1$^{\prime}$ of the galaxy cluster center for optical, mid-infrared (MIR), radio, and optical-to-MIR color-selected AGNs. \cite{Mo18} also find that the radio-selected AGN fraction is more than 2.5 times that of the field, implying that the centers of clusters are conducive to triggering radio emission in AGNs.

With this broader context in mind, \cite{EM19} conducted a pilot study of the morphology of 10 of the most extended, centrally located radio sources in MaDCoWS clusters. \cite{EM19} find a tight relationship between the size of the radio source and the distance from the cluster center. In this present work, we present an analysis of the full sample of 52 central extended radio sources within 1$^{\prime}$ of \madcows\ cluster centers. We determine the morphological properties of the extended radio sources and investigate correlations between radio source properties, host-galaxy properties, and cluster properties. In Section \ref{sample}, we describe sample selection and observations. In Section \ref{sect:radio}, we classify the radio morphology of the sources. In Section \ref{sect:hg}, we discuss the host-galaxy properties. Lastly, in Section \ref{sect:environ}, we investigate and discuss correlations between the radio source properties and properties of the environment. Throughout this work, we adopt the flat $\Lambda$CDM cosmological model with a \cite{Planck15} cosmology, $H_0$ = 67.8 km s$^{-1}$, $\Omega_{m}$ = 0.308, $\Omega_{\Lambda}$ = 0.692, and $n_s$ = 0.968. Unless otherwise noted, \spz\ magnitudes are on the Vega system and Pan-STARRS magnitudes are on the AB system in order to remain on the fiducial system of the surveys\footnote{The conversion from Vega to AB for [3.6] is [3.6]$_{\mathrm{Vega}}$ = [3.6]$_{\mathrm{AB}}$ - 2.79 and for [4.5] the conversion is [4.5]$_{\mathrm{Vega}}$ = [4.5]$_{\mathrm{AB}}$ - 3.26, assuming z=1.}.

\section{Sample Selection and Observations} \label{sample}
\begin{deluxetable*}{cccccc}
	\tablecaption{Cluster Properties and VLA Observations\label{tb:vla_obs}}
	\tablehead{\colhead{Cluster} & \colhead{$z_{\rm{phot}}$} & \colhead{$\lambda_{\rm{15}}$} & \colhead{Band-Config.-} & \colhead{Beam} & \colhead{RMS} \\ 
	& & N & \colhead{Obs. Sem.} & \colhead{\as}& \colhead{($\mu$Jy/beam)}}
	\startdata
        MOO J0015+0801 & 0.9$_{-0.04}^{+0.04}$ & 59$\pm$8 & L-A-16B & 1.7$^{\prime\prime}\times$1.2$^{\prime\prime}$ & 45.0 \\
MOO J0100$-$0151 & 1.1$_{-0.05}^{+0.06}$ & 37$\pm$7 & L-A-18A & 2.1$^{\prime\prime}\times$1.2$^{\prime\prime}$ & 50.0 \\
MOO J0121$-$0145 & 0.98$_{-0.07}^{+0.07}$ & 38$\pm$7 & L-A-16B & 1.1$^{\prime\prime}\times$1.0$^{\prime\prime}$ & 24.0 \\
MOO J0228$-$0644 & 0.86$_{-0.06}^{+0.09}$ & 22$\pm$6 & L-A-16B & 2.5$^{\prime\prime}\times$1.1$^{\prime\prime}$ & 30.0 \\
MOO J0231$-$0212 & 1.43$_{-0.07}^{+0.08}$ & 23$\pm$5 & C-B-17B & 1.9$^{\prime\prime}\times$1.0$^{\prime\prime}$ & 8.35 \\
MOO J0250$-$0443 & $-$* & $-$ & L-A-16B & 1.7$^{\prime\prime}\times$1.1$^{\prime\prime}$ & 20.0 \\
MOO J0300+0124 & 1.33$_{-0.06}^{+0.04}$ & 37$\pm$6 & L-A-16B & 1.4$^{\prime\prime}\times$1.1$^{\prime\prime}$ & 33.0 \\
MOO J0900+0407 & 0.84$_{-0.05}^{+0.07}$ & 14$\pm$5 & C-B-17B & 2.8$^{\prime\prime}\times$1.0$^{\prime\prime}$ & 8.7 \\
MOO J0901+3341 & 1.02$_{-0.07}^{+0.08}$ & 27$\pm$6 & C-B-17B & 2.0$^{\prime\prime}\times$1.1$^{\prime\prime}$ & 6.7 \\
MOO J0903+1319 & $-$* & $-$ & C-B-17B & 1.5$^{\prime\prime}\times$1.1$^{\prime\prime}$ & 7.0 \\
MOO J0917+1456 & 1.05$_{-0.07}^{+0.07}$ & 30$\pm$6 & C-B-17B & 2.0$^{\prime\prime}\times$1.1$^{\prime\prime}$ & 7.3 \\
MOO J0920$-$0755 & 1.1$_{-0.06}^{+0.06}$ & 31$\pm$6 & C-B-17B & 2.4$^{\prime\prime}\times$1.0$^{\prime\prime}$ & 7.8 \\
MOO J0928$-$0126 & $-$* & $-$ & C-B-17B & 3.0$^{\prime\prime}\times$1.0$^{\prime\prime}$ & 7.2 \\
MOO J0959+3903 & $-$* & $-$ & C-B-17B & 1.8$^{\prime\prime}\times$1.0$^{\prime\prime}$ & 7.0 \\
MOO J1004$-$0257 & 0.98$_{-0.07}^{+0.07}$ & 32$\pm$6 & L-A-18A & 2.4$^{\prime\prime}\times$1.1$^{\prime\prime}$ & 30.0 \\
MOO J1007+0139 & 0.94$_{-0.06}^{+0.06}$ & 28$\pm$6 & C-B-17B & 1.2$^{\prime\prime}\times$1.0$^{\prime\prime}$ & 7.2 \\
MOO J1011+3919 & $-$* & $-$ & C-BnA-17B & 2.4$^{\prime\prime}\times$0.5$^{\prime\prime}$ & 7.6 \\
MOO J1012+0843 & 1.14$_{-0.08}^{+0.1}$ & 25$\pm$5 & L-A-18A & 1.8$^{\prime\prime}\times$1.2$^{\prime\prime}$ & 32.0 \\
MOO J1020+1231 & 0.89$_{-0.07}^{+0.09}$ & 13$\pm$5 & L-A-18A & 2.1$^{\prime\prime}\times$1.3$^{\prime\prime}$ & 30.0 \\
MOO J1020+2412 & 1.01$_{-0.06}^{+0.06}$ & 25$\pm$6 & L-A-18A & 1.9$^{\prime\prime}\times$1.3$^{\prime\prime}$ & 32.0 \\
MOO J1021+1800 & $-$* & $-$ & C-B-17B & 1.4$^{\prime\prime}\times$1.2$^{\prime\prime}$ & 6.7 \\
MOO J1034+3104 & $-$* & $-$ & L-A-18A & 1.5$^{\prime\prime}\times$1.2$^{\prime\prime}$ & 40.0 \\
MOO J1037+0433 & $-$* & $-$ & L-A-18A & 2.2$^{\prime\prime}\times$1.2$^{\prime\prime}$ & 34.0 \\
MOO J1053+1052 & 0.99$_{-0.06}^{+0.07}$ & 44$\pm$7 & L-A-18A & 1.9$^{\prime\prime}\times$1.1$^{\prime\prime}$ & 14.0 \\
MOO J1117+1514 & 1.13$_{-0.09}^{+0.1}$ & 25$\pm$5 & C-BnA-17B & 2.4$^{\prime\prime}\times$0.4$^{\prime\prime}$ & 7.6 \\
MOO J1138+5031 & 1.29$_{-0.1}^{+0.07}$ & 23$\pm$5 & C-B-17B & 2.9$^{\prime\prime}\times$1.0$^{\prime\prime}$ & 6.8 \\
MOO J1140+4454 & 1.09$_{-0.04}^{+0.04}$ & 36$\pm$6 & C-B-17B & 2.7$^{\prime\prime}\times$1.0$^{\prime\prime}$ & 7.9 \\
MOO J1215$-$0630 & 1.02$_{-0.06}^{+0.07}$ & 27$\pm$6 & L-A-18A & $-$ & $-$ \\
MOO J1238$-$0232 & 0.94$_{-0.09}^{+0.1}$ & 13$\pm$5 & L-A-18A & 1.5$^{\prime\prime}\times$1.1$^{\prime\prime}$ & 13.0 \\
MOO J1252+0700 & 0.94$_{-0.08}^{+0.09}$ & 28$\pm$6 & L-A-18A & 1.3$^{\prime\prime}\times$1.2$^{\prime\prime}$ & 12.0 \\
MOO J1308+1622 & 1.02$_{-0.09}^{+0.09}$ & 18$\pm$5 & C-B-17B & 1.1$^{\prime\prime}\times$1.1$^{\prime\prime}$ & 7.0 \\
MOO J1312+2002 & 0.94$_{-0.07}^{+0.07}$ & 26$\pm$6 & C-B-17B & 2.4$^{\prime\prime}\times$1.1$^{\prime\prime}$ & 7.4 \\
MOO J1319+5519 & 0.94$_{-0.04}^{+0.05}$ & 44$\pm$7 & C-B-17B & 1.8$^{\prime\prime}\times$1.0$^{\prime\prime}$ & 7.1 \\
MOO J1334+1443 & 0.98$_{-0.1}^{+0.11}$ & 21$\pm$6 & C-B-17B & 1.2$^{\prime\prime}\times$1.0$^{\prime\prime}$ & 6.2 \\
MOO J1353$-$0541 & 0.94$_{-0.07}^{+0.07}$ & 22$\pm$5 & L-A-18A & 1.6$^{\prime\prime}\times$1.1$^{\prime\prime}$ & 33.0 \\
MOO J1355+6116 & 0.38$_{-0.05}^{+0.03}$ & 51$\pm$8 & C-B-17B & 1.3$^{\prime\prime}\times$1.0$^{\prime\prime}$ & 6.5 \\
MOO J1358+2158 & 0.99$_{-0.06}^{+0.07}$ & 39$\pm$7 & L-A-16B & 1.1$^{\prime\prime}\times$1.0$^{\prime\prime}$ & 31.0 \\
MOO J1401+0750 & 1.08$_{-0.08}^{+0.08}$ & 39$\pm$7 & C-B-17B & 1.2$^{\prime\prime}\times$1.0$^{\prime\prime}$ & 6.75 \\
MOO J1412+4846 & $-$* & $-$ & L-A-16B & 3.5$^{\prime\prime}\times$1.0$^{\prime\prime}$ & 30.0 \\
MOO J1435+4759 & 1.02$_{-0.05}^{+0.06}$ & 44$\pm$7 & L-A-16B & 3.5$^{\prime\prime}\times$1.0$^{\prime\prime}$ & 27.0 \\
MOO J1440+2628 & $-$* & $-$ & C-B-17B & 1.3$^{\prime\prime}\times$1.1$^{\prime\prime}$ & 6.45 \\
MOO J1506+5137 & 1.09$_{-0.03}^{+0.03}$ & 75$\pm$9 & C-B-17B & 1.0$^{\prime\prime}\times$0.8$^{\prime\prime}$ & 8.0 \\
MOO J1507+3126 & 0.88$_{-0.04}^{+0.03}$ & 40$\pm$7 & C-B-17B & 1.6$^{\prime\prime}\times$1.0$^{\prime\prime}$ & 6.7 \\
MOO J1516$-$0435 & $-$* & $-$ & L-A-18A & 1.5$^{\prime\prime}\times$1.1$^{\prime\prime}$ & 33.0 \\
MOO J1555+3259 & 0.98$_{-0.07}^{+0.07}$ & 24$\pm$6 & C-BnA-17B & 1.1$^{\prime\prime}\times$0.4$^{\prime\prime}$ & 5.9 \\
MOO J1619+5323 & 1.21$_{-0.09}^{+0.08}$ & 25$\pm$5 & L-A-18A & 1.5$^{\prime\prime}\times$1.0$^{\prime\prime}$ & 33.0 \\
MOO J1622+1324 & $-$* & $-$ & C-B-17B & 1.1$^{\prime\prime}\times$1.0$^{\prime\prime}$ & 6.9 \\
MOO J1634+4021 & 1.16$_{-0.02}^{+0.02}$ & 38$\pm$6 & C-B-17B & 1.7$^{\prime\prime}\times$1.1$^{\prime\prime}$ & 6.0 \\
MOO J1731+5857 & 1.02$_{-0.08}^{+0.08}$ & 29$\pm$6 & L-A-16B & 1.4$^{\prime\prime}\times$1.0$^{\prime\prime}$ & 19.0 \\
MOO J2247+0507 & 1.02$_{-0.05}^{+0.05}$ & 21$\pm$5 & L-A-16B & 1.4$^{\prime\prime}\times$1.1$^{\prime\prime}$ & 49.0 \\
MOO J2306+0951 & 1.09$_{-0.11}^{+0.12}$ & 20$\pm$5 & L-A-18A & $-$ & $-$ \\
MOO J2321+0119 & $-$* & $-$ & C-B-17B & 1.6$^{\prime\prime}\times$1.3$^{\prime\prime}$ & 6.0 \\
	\enddata
	\tablecomments{The prefix of the cluster name stands for Massive Overdense Object. Column 2: the photometric redshift. $^*$We note that some clusters lack IRAC data for the calculation of a photometric redshift, thus we assume the median redshift of the MaDCoWS survey for these clusters \citep[$z=1.06$,][]{Gonzalez19}. Column 3: richness of the cluster defined in \S\ref{sect:spz}. Those that do not have \spz\ data do not have a richness. Column 4: the VLA band, configuration, and observing semester of the observation. Column 5: the beam size of each observation. Column 6: the RMS of the VLA image. MOO J1215$-$0630 and MOO J2306+0951 had unusable datasets.}
\end{deluxetable*}

We construct our sample by cross-matching two surveys: \madcows\ and the Faint Images of the Radio Sky at Twenty-cm \citep[FIRST,][]{Becker94,Becker95} Survey. The primary \madcows\ search covers 17,668 deg$^2$ of the extragalactic sky at $\delta$ $>$ $-$30$^{\circ}$ using a combination of data from the \textit{Wide-field Infrared Survey Explorer} \citep[\textit{WISE};][]{Wright10} and the Panoramic Survey Telescope and Rapid Response System \citep[Pan-STARRS;][]{PS16} to detect cluster candidates in the redshift range of 0.7 $\lesssim z \lesssim$ 1.5. \cite{Gonzalez19} report the 2681 highest amplitude detections. The cluster centers are defined by the catalog coordinates from the original \textit{WISE}---PanSTARRS search as described in \cite{Gonzalez19} and have a positional uncertainty of 21\as (see \citealt{Gonzalez19} for a more detailed discussion).

\cite{Mo18} find that 19\% of \madcows\ clusters within the FIRST footprint have at least one FIRST source coincident with the inner $1\arcmin$ region compared to the 5\% found if the center is offset to a random position far from the cluster. From the 1300 highest significance \madcows\ clusters in the FIRST footprint, we identified a parent sample of \madcows\ clusters with the most extended radio sources near the cluster center. These clusters satisfy the criteria of having FIRST sources with deconvolved sizes exceeding 6.5$^{\prime\prime}$ (50 kpc at $z \simeq 1$) within 1$^{\prime}$ of the cluster center. The threshold of 6.5$^{\prime\prime}$ was chosen to ensure the sources were truly extended by choosing a size threshold 30\% larger than the 5\as\ resolution of FIRST. 51 clusters met these criterion and were the primary sources targeted for VLA follow-up observations, the first ten of which were observed in the 2016B observing semester and are presented in \cite{EM19}. Additionally, an extra cluster MOO J1319+5519 that contains a radio source that is slightly smaller than the 6.5$^{\prime\prime}$ threshold was included in the VLA observational program. This gives a full sample of 52 sources. In some cases, there were additional FIRST sources within $\sim$ 1$^{\prime}$ of the cluster center that are less extended and we refer to these sources as the secondary sources. The analysis here is limited to the primary sources. We note that the most extended and diffuse sources could be missing from the sample due to the pre-selection based on FIRST.

Since there are no spectroscopic redshifts available to confirm cluster membership, there is a possibility that the extended sources within 1$^{\prime}$ of the cluster center are chance superpositions. We consider the interloper fraction in more detail in \S\ref{sect:interlopers}. 

\subsection{FIRST Data}
The FIRST survey covers 10,575 deg$^2$ of the North and South Galactic Caps. These data were taken using the VLA in the B-configuration in L-band. FIRST has a typical RMS of 0.15 mJy and a resolution of 5$^{\prime\prime}$. Table \ref{tb:morph} lists relevant properties obtained from the FIRST catalog.

We calculate the power (radio luminosity) of the FIRST sources,
\begin{equation}\label{eqn:power}
    P_{1.4} = 4\pi{D_L}^2S_{1.4}(1+z)^{\alpha-1},
\end{equation}
where $D_L$ is the luminosity distance at the redshift of the cluster, $S_{1.4}$ is the integrated radio flux at 1.4 GHz from FIRST, (1 + $z$)$^{\alpha-1}$ includes both the distance dimming and K-correction, and $\alpha$ is the radio spectral index ($S_{\nu} \propto \nu^{-\alpha}$). Typical values of $\alpha$ for extended radio sources range from 0.7 to 0.8 \citep[]{Kellermann88,Condon92,PetersonBook,L&M07,Miley08,Tiwari16} and we adopt $\alpha$=0.8 as in \citet[]{Chiaberge09}, \citet[]{Gralla11}, and \citet[]{Yuan16}. We assume the radio source is at the photometric redshift of the cluster (see Table \ref{tb:vla_obs} and \S\ref{sect:spz}). 

We note that at $z$ = 1, a 1 mJy (the FIRST flux limit) source corresponds to P$_{1.4}$ = 4.7$\times 10^{24}$ W Hz$^{-1}$. This limit is barely below the frequently quoted power boundary between the two FR classes \citep[P$_{1.4}$$\sim$ 10$^{25}$ W Hz$^{-1}$;][]{FR74,LO96}. Thus, any FR I sources we detect will be at the luminous end of the FR I distribution. 

\begin{figure*}
    \centering
	\includegraphics[width=\linewidth, keepaspectratio]{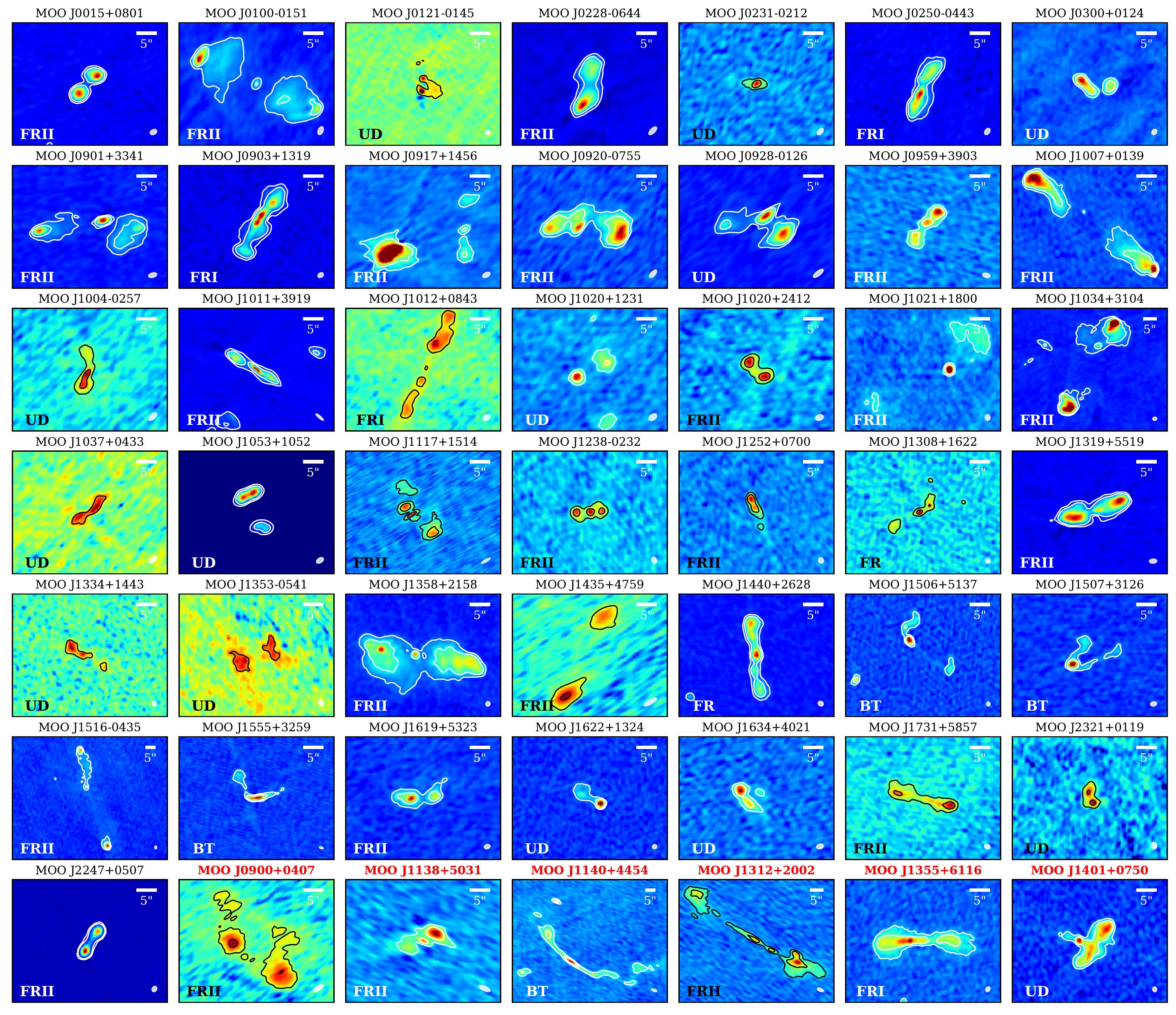}
	\caption{VLA images of the targeted radio source(s) in each cluster. Most images are 30$^{\prime\prime}$ $\times$ 30$^{\prime\prime}$ (250 kpc $\times$ 250 kpc) indicated by the scale bar in the upper right hand corner. North is up and east is to the left. Each image is scaled by the square root of the flux distribution. The contours levels are 4$\sigma$ and 16$\sigma$, except for MOO J0228$-$0644, MOO J1011+3919, and MOO J2247+2247 which are 8$\sigma$ and 32$\sigma$ to showcase the morphology of the source. The synthesized beam size is shown in the lower right hand corner. The black and white contours were chosen to guide the eye. The morphology classification is indicated in the lower left hand corner (see \S\ref{classification}). Those that have red, bold titles (most of the last row of the figure) are identified as low redshift interlopers in \S\ref{sect:interlopers}. We find a majority of the sources to be FR II sources.}
	\label{fig:radio_sources}
\end{figure*}

\subsection{VLA Follow-up Observations and Image Processing}\label{sect:vla}
VLA high-resolution follow-up observations were taken over three observing semesters: 16B (PI: Gonzalez, 16B-289), 17B (PI: Gonzalez, 17B-197), and 18A (PI: Moravec, 18A-039). See Table \ref{tb:vla_obs} for details. The results of the 10 clusters observed in 16B are published in \cite{EM19}, but are also included in this work. 

The data taken in 16B and 18A were taken in L-band in the A configuration. The observations were centered at 1.4 GHz (21cm) with a bandwidth of 600 MHz. Given this configuration and band, the resolution was $\sim$1.3$^{\prime\prime}$ using robust weighting (see Table \ref{tb:vla_obs} for exact beam sizes). The primary beam was 30$^{\prime}$ and the largest angular scale which we can detect is 36.\as\ During 16B, each target was observed twice for nine and a half minutes within a scheduling block (a total of $\sim$19 minutes on source) to reach the desired sensitivity. Between 16B and 18A, updated L-band system equivalent flux density (SEFD) measurements were integrated into the exposure time calculator and thus we required less integration time in 18A to reach the equivalent sensitivity that was achieved in 16B. During 18A, each target was observed twice for six minutes within a scheduling block for a total of $\sim$12 minutes on source.

The data taken in 17B were taken in C-band in the B configuration. The observations were centered at 5.5 GHz with a bandwidth of 1.9 GHz. Given this configuration and band, the resolution was $\sim$1.0$^{\prime\prime}$ using robust weighting (see Table\ref{tb:vla_obs} for exact beam sizes). The primary beam was 8$^{\prime}$ and the largest angular scale which we can detect is 29.\as\ During 17B, each target was observed three times for nine minutes each within a scheduling block for a total of $\sim$28 minutes on source. A few cluster C-band datasets were taken in the BnA hybrid configuration during the transition from B configuration to A configuration in early February 2018 where the northern arm is in the extended A configuration and eastern and western arms are in B configuration. The resolution achieved with this configuration is similar to those achieved in B configuration. For all observations the correlator was configured with 16 spectral windows, each with 64 channels. 

Most of the data were flagged, calibrated, and imaged with Common Astronomy Software Applications (CASA) version 5.0 or later \citep[]{casa}. All measurement sets were first processed through the VLA CASA Calibration Pipeline for basic flagging and calibration. We created images of each cluster by applying the TCLEAN algorithm. We cleaned the L-band images to an average depth of $\sim$28 $\mu$Jy per beam and $\sim$7 $\mu$Jy per beam for the C-band images.

The RMS ($\mu$Jy/beam) value for each image was determined by the following process. If possible, four rectangular regions that were near the targeted source, but did not encompass the source were drawn in the viewer. These regions were drawn in such a way as to be free of detected radio sources. Then, these individual RMS measurements were averaged together to produce the final RMS which is reported in Table \ref{tb:vla_obs} for each image. 

A pixel scale of 0.28$^{\prime\prime}$ for L-band images and 0.24$^{\prime\prime}$ for C-band images, specmode=`mfs', and WEIGHTING = BRIGGS (robust=0.5) were used for all images. Analysis is restricted to the inner 3$^{\prime}$ of the primary beam and in most cases to the inner 1$^{\prime}$. Thus, primary beam correction was not necessary. In many cases, we performed several rounds of phase-only self-calibration to increase the S/N ratio (typically by a factor of 3), reduce the prominence of improper cleaning artifacts, and recover more of the source structure.

Two clusters (MOO J1215$-$0630 and MOO J2306+0951) had unusable data. MOO J1215$-$0630 had a bright source in the upper right of the field that could not be cleaned well enough to allow the source of interest to be distinguished. For MOO J2306+0951, an extended source was unintentionally used for calibration which made imaging difficult. Due to these data quality issues, the final sample is reduced from 52 to 50.

\subsection{Spitzer Observations} \label{sect:spz}
A total of 36 of the 50 clusters presented in this work were also observed with \textit{Spitzer} (90177 and 11080, PI: Gonzalez). Each cluster was observed in each $[3.6]$ and $[4.5]$ for a total of $\sim$180 seconds using a set of $6 \times 30$s exposures, with a medium scale cycling dither pattern.

The catalogs were generated using a procedure similar to that of \cite{Wylezalek13} in which the Corrected Basic Calibrated Data (cBCD) was reduced and mosaicked using IRACproc (\citealt{Schuster06}), the MOPEX package (\citealt{MK05}), and a resampled pixel scale of 0\farcs6. The MOPEX outlier (e.g., cosmic ray, bad pixel) rejection was optimized for the regions of deepest coverage in the center of the maps corresponding to the position of the \madcows\ detection. Photometry was performed using SExtractor \citep[]{BA96} with an aperture diameter of 4$^{\prime\prime}$, corrected to total. The catalogs reach a uniform limit of 10 $\mu$Jy in [3.6] and [4.5]. 

The photometric redshifts are listed in Table \ref{tb:vla_obs} and are derived from the [3.6]$-$[4.5] and Pan-STARRS $i-$[3.6] colors of galaxies within 1$^{\prime}$ of the cluster location, which are compared with a Flexible Stellar Population Synthesis (FSPS) model \citep{c09,c10}. Details can be found in \cite{Gonzalez19}. For those clusters that lack IRAC data, we assume the median photometric redshift of the \madcows\ survey \citep[$z=1.06$,][]{Gonzalez19}.

For the clusters that have \spz\ data the richness is defined as $\lambda_{15}$ = N$-$N$_{\mathrm{field}}$, where N is the total number of color-selected galaxies within the metric aperture that have fluxes that satisfy f$_{4.5}>~$15$\mu$Jy. Details can be found in \cite{Gonzalez19}. The richnesses are recorded in Table \ref{tb:vla_obs}.

While the primary cluster coordinates are from the original survey \citep{Gonzalez19}, the \spz\ photometry provides an independent measurement of the cluster center in a similar manner. We use the average distance from the \wise\ and \spz\ center when possible (see \S\ref{sect:LAS_v_D}). Some clusters do not have a robust \textit{Spitzer} center and in these cases the \wise\ center is adopted as the center.

\begin{deluxetable*}{cllcccccc}
    \tablecolumns{9}
	\tablecaption{Radio Source Properties\label{tb:morph}}
	\tablehead{
	\colhead{Host Cluster} & \colhead{Radio Source} & \colhead{Radio Source} & \colhead{$R_{\mathrm{cc}}$} & \colhead{FIRST Int. Flux} & \colhead{$P_{1.4}$} & \colhead{Morph.} & \colhead{LAS} & \colhead{Linear Size} \\ 
	\colhead{} & \colhead{RA} & \colhead{Dec} & \colhead{($^{\prime\prime}$)} & \colhead{(mJy)} & \colhead{(10$^{26}$ W Hz$^{-1}$)} & \colhead{} & \colhead{($^{\prime\prime}$)} & \colhead{(kpc)}}
	\startdata
        MOO J0015+0801 & 00:15:24.50 & +08:01:15.3 & 21.9$\pm$2.1 & 49.5$\pm$0.1 & 1.85 & FRII & 10.7$\pm$0.9 & 85$\pm$7 \\
MOO J0100$-$0151 & 01:01:02.29 & $-$01:51:20.7 & 56.3$\pm$2.8 & 60.1$\pm$0.4 & 3.62 & FRII & 34.4$\pm$1.0 & 288$\pm$8 \\
MOO J0228$-$0644 & 02:28:02.27 & $-$06:44:37.1 & 16.0$\pm$4.8 & 61.2$\pm$0.1 & 2.06 & FRII & 16.2$\pm$1.2 & 128$\pm$9 \\
MOO J0231$-$0212 & 02:31:13.07 & $-$02:12:12.1 & 18.1$\pm$0.4 & 2.4$\pm$0.1 & 0.27 & UD & 6.3$\pm$0.9 & 54$\pm$7 \\
MOO J0250$-$0443 & 02:50:44.52 & $-$04:43:02.2 & 19.0$\pm$4.8 & 22.9$\pm$0.1 & 1.26 & FRI & 16.5$\pm$0.9 & 137$\pm$7 \\
MOO J0300+0124 & 03:00:12.93 & +01:24:59.6 & 17.0$\pm$9.0 & 10.7$\pm$0.1 & 1.01 & UD & 7.4$\pm$0.7 & 63$\pm$6 \\
MOO J0300+0124 & 03:00:12.52 & +01:24:59.2 & 12.1$\pm$7.7 & 10.7$\pm$0.1 & 1.01 & UD & 4.3$\pm$0.7 & 37$\pm$6 \\
MOO J0900+0407 & 09:00:47.87 & +04:07:56.4 & 55.7$\pm$2.1 & 18.5$\pm$0.5 & 0.59 & FRII & 22.4$\pm$1.4 & $-$ \\
MOO J0901+3341 & 09:01:02.43 & +33:40:52.2 & 72.9$\pm$11.0 & 26.0$\pm$0.3 & 1.31 & FRII & 28.7$\pm$1.0 & 237$\pm$8 \\
MOO J0903+1319 & 09:03:04.96 & +13:19:51.7 & 8.4$\pm$4.8 & 65.2$\pm$0.1 & 3.60 & FRI & 20.8$\pm$0.7 & 173$\pm$5 \\
MOO J0917+1456 & 09:17:20.12 & +14:56:10.6 & 42.0$\pm$6.5 & 4.2$\pm$0.1 & 0.23 & FRII & 16.8$\pm$1.0 & 139$\pm$8 \\
MOO J0920$-$0755 & 09:20:53.97 & $-$07:55:30.5 & 12.7$\pm$2.4 & 28.3$\pm$0.2 & 1.70 & FRII & 22.4$\pm$1.2 & 188$\pm$10 \\
MOO J0928$-$0126 & 09:28:08.73 & $-$01:25:38.0 & 26.5$\pm$4.8 & 18.5$\pm$0.3 & 1.02 & UD & 20.6$\pm$1.5 & 171$\pm$12 \\
MOO J0959+3903 & 09:59:57.75 & +39:03:49.0 & 1.7$\pm$4.8 & 5.2$\pm$0.1 & 0.29 & FRII & 12.17$\pm$0.9 & 101$\pm$7 \\
MOO J1004$-$0257 & 10:04:38.97 & $-$02:57:45.8 & 23.0$\pm$4.5 & 2.8$\pm$0.1 & 0.13 & UD & 12.1$\pm$1.2 & 99$\pm$9 \\
MOO J1007+0139 & 10:07:21.70 & +01:39:26.3 & 31.6$\pm$4.8 & 54.9$\pm$0.2 & 2.28 & FRII & 40.4$\pm$0.6 & 327$\pm$4 \\
MOO J1011+3919 & 10:11:55.22 & +39:20:04.1 & 30.8$\pm$4.8 & 47.2$\pm$0.1 & 2.60 & FRII & 16.0$\pm$1.2 & 133$\pm$10 \\
MOO J1012+0843 & 10:12:17.22 & +08:42:26.3 & 62.0$\pm$8.2 & 5.0$\pm$0.1 & 0.33 & FRI & 28.4$\pm$0.9 & 240$\pm$7 \\
MOO J1020+1231 & 10:20:40.09 & +12:31:45.7 & 19.8$\pm$5.1 & 3.6$\pm$0.1 & 0.13 & UD & 5.9$\pm$1.0 & 47$\pm$7 \\
MOO J1020+2412 & 10:20:04.89 & +24:12:13.4 & 11.6$\pm$7.4 & 5.2$\pm$0.1 & 0.26 & FRII & 8.4$\pm$1.0 & 69$\pm$8 \\
MOO J1021+1800 & 10:21:54.93 & +18:00:54.6 & 4.6$\pm$4.8 & 10.1$\pm$0.3 & 0.56 & FRII & 34.6$\pm$0.7 & 288$\pm$5 \\
MOO J1034+3104 & 10:34:32.73 & +31:03:52.1 & 22.6$\pm$4.8 & 281.7$\pm$0.2 & 15.50 & FRII & 40.4$\pm$0.8 & 336$\pm$6 \\
MOO J1037+0433 & 10:37:11.81 & +04:33:07.7 & 16.5$\pm$4.8 & 2.4$\pm$0.1 & 0.13 & UD & 10.8$\pm$1.1 & 90$\pm$9 \\
MOO J1053+1052 & 10:53:17.73 & +10:52:29.5 & 59.2$\pm$3.9 & 2.5$\pm$0.1 & 0.12 & UD & 5.1$\pm$0.9 & 41$\pm$7 \\
MOO J1117+1514 & 11:17:36.53 & +15:14:44.3 & 13.9$\pm$4.8 & 13.5$\pm$0.2 & 0.87 & FRII & 12.0$\pm$1.2 & 101$\pm$10 \\
MOO J1138+5031 & 11:38:14.25 & +50:31:45.4 & 24.5$\pm$1.0 & 4.9$\pm$0.2 & 0.43 & FRII & 11.8$\pm$1.4 & $-$ \\
MOO J1140+4454 & 11:40:23.86 & +44:54:17.7 & 21.3$\pm$5.1 & 13.6$\pm$0.6 & 0.80 & BT & 61.5$\pm$1.4 & $-$ \\
MOO J1238$-$0232 & 12:38:57.75 & $-$02:31:41.5 & 35.7$\pm$0.1 & 1.7$\pm$0.1 & 0.07 & FRII & 9.2$\pm$0.8 & 74$\pm$6 \\
MOO J1252+0700 & 12:52:46.56 & +07:00:54.2 & 11.6$\pm$1.2 & 2.1$\pm$0.1 & 0.09 & FRII & 7.0$\pm$0.7 & 56$\pm$5 \\
MOO J1308+1622 & 13:08:24.27 & +16:22:51.4 & 21.4$\pm$9.4 & 4.0$\pm$0.2 & 0.20 & FRI/II & 14.3$\pm$0.6 & 118$\pm$4 \\
MOO J1312+2002 & 13:12:49.56 & +20:02:26.1 & 18.2$\pm$4.8 & 14.8$\pm$0.2 & 0.61 & FRII & 60.2$\pm$1.2 & $-$ \\
MOO J1319+5519 & 13:19:41.94 & +55:19:01.8 & 14.9$\pm$3.4 & 117.9$\pm$0.2 & 4.89 & FRII & 18.4$\pm$0.9 & 149$\pm$7 \\
MOO J1334+1443 & 13:34:46.43 & +14:42:58.6 & 34.2$\pm$18.0 & 3.2$\pm$0.1 & 0.15 & UD & 11.6$\pm$0.6 & 95$\pm$4 \\
MOO J1353$-$0541 & 13:53:46.94 & $-$05:41:23.5 & 33.7$\pm$3.8 & 5.0$\pm$0.2 & 0.21 & UD & 12.8$\pm$0.8 & 103$\pm$6 \\
MOO J1355+6116 & 13:55:56.60 & +61:16:48.1 & 43.8$\pm$4.6 & 28.1$\pm$0.3 & 0.14 & FRI & 25.0$\pm$0.6 & $-$ \\
MOO J1358+2158 & 13:58:22.94 & +21:59:19.6 & 43.7$\pm$4.7 & 102.1$\pm$0.4 & 4.79 & FRII & 31.6$\pm$0.6 & 259$\pm$4 \\
MOO J1401+0750 & 14:01:22.23 & +07:51:05.3 & 26.5$\pm$4.8 & 20.4$\pm$0.1 & 1.18 & UD & 14.8$\pm$0.6 & $-$ \\
MOO J1412+4846 & 14:12:58.06 & +48:47:01.2 & 52.6$\pm$4.8 & 10.4$\pm$0.4 & 0.57 & FRII & 52.5$\pm$1.8 & 437$\pm$15 \\
MOO J1435+4759 & 14:35:19.41 & +48:00:02.5 & 47.6$\pm$1.4 & 6.5$\pm$0.3 & 0.33 & FRII & 29.0$\pm$1.8 & 239$\pm$14 \\
MOO J1440+2628 & 14:40:09.82 & +26:28:55.9 & 6.5$\pm$4.8 & 24.7$\pm$0.2 & 1.36 & FRI/II & 21.0$\pm$0.7 & 175$\pm$5 \\
MOO J1506+5137 & 15:06:20.40 & +51:36:53.7 & 11.4$\pm$3.6 & 12.2$\pm$0.2 & 0.72 & BT & 18.0$\pm$0.5 & 150$\pm$4 \\
MOO J1507+3126 & 15:07:53.03 & +31:27:08.6 & 27.3$\pm$4.8 & 5.7$\pm$0.2 & 0.20 & BT & 14.5$\pm$0.8 & 115$\pm$6 \\
MOO J1516$-$0435 & 15:16:40.11 & $-$04:35:39.5 & 16.8$\pm$4.8 & 21.2$\pm$0.4 & 1.17 & FRII & 52.6$\pm$0.7 & 438$\pm$5 \\
MOO J1555+3259 & 15:55:23.06 & +32:59:56.5 & 27.9$\pm$6.0 & 12.9$\pm$0.1 & 0.59 & BT & 8.9$\pm$0.5 & 72$\pm$4 \\
MOO J1619+5323 & 16:19:31.64 & +53:23:09.6 & 27.7$\pm$4.8 & 15.3$\pm$0.1 & 1.15 & FRII & 14.3$\pm$0.7 & 121$\pm$5 \\
MOO J1622+1324 & 16:22:26.77 & +13:24:03.8 & 44.2$\pm$4.8 & 3.6$\pm$0.1 & 0.20 & UD & 8.9$\pm$0.6 & 74$\pm$5 \\
MOO J1634+4021 & 16:34:36.42 & +40:21:31.6 & 7.9$\pm$4.8 & 3.6$\pm$0.1 & 0.25 & UD & 10.1$\pm$0.8 & 85$\pm$6 \\
MOO J1731+5857 & 17:31:00.30 & +58:58:05.7 & 38.1$\pm$0.4 & 5.6$\pm$0.2 & 0.28 & FRII & 17.4$\pm$0.7 & 143$\pm$5 \\
MOO J2247+0507 & 22:47:15.27 & +05:08:07.2 & 12.3$\pm$4.2 & 332.6$\pm$0.1 & 16.70 & FRII & 9.9$\pm$0.7 & 81$\pm$5 \\
MOO J2321+0119 & 23:21:07.40 & +01:19:14.0 & 41.3$\pm$4.8 & 2.9$\pm$0.1 & 0.16 & UD & 6.7$\pm$0.8 & 55$\pm$6 \\
	\enddata
	\tablecomments{We omit MOO J1215$-$0630 and MOO J2306+0019 due to data quality issues that made morphological classification impossible (see \S\ref{sect:vla}) and MOO J0121$-$0145 because it has an ambiguous morphology (see \S\ref{LTS}). Columns 2 \& 3: The RA and Dec of the inferred origin of the radio emission based on the morphological classification. Column 4: The distance of the source from the cluster center. Column 5: The FIRST integrated flux. Column 6: The power calculated using Eqn.~\ref{eqn:power} and the FIRST integrated flux. Column 7: The VLA follow-up morphology (see \S\ref{sect:radio}). Column 8: The largest angular size (LAS) as defined in \S\ref{extent}. Column 9: The linear size is the LAS converted to kpc assuming the radio source is at the redshift of the cluster. If the source is identified as an interloper per \S\ref{sect:interlopers}, we cannot accurately calculate the linear size; it is thus indicated by a --.} 
\end{deluxetable*}

\begin{figure}
    \centering
	\includegraphics[width=\linewidth, keepaspectratio]{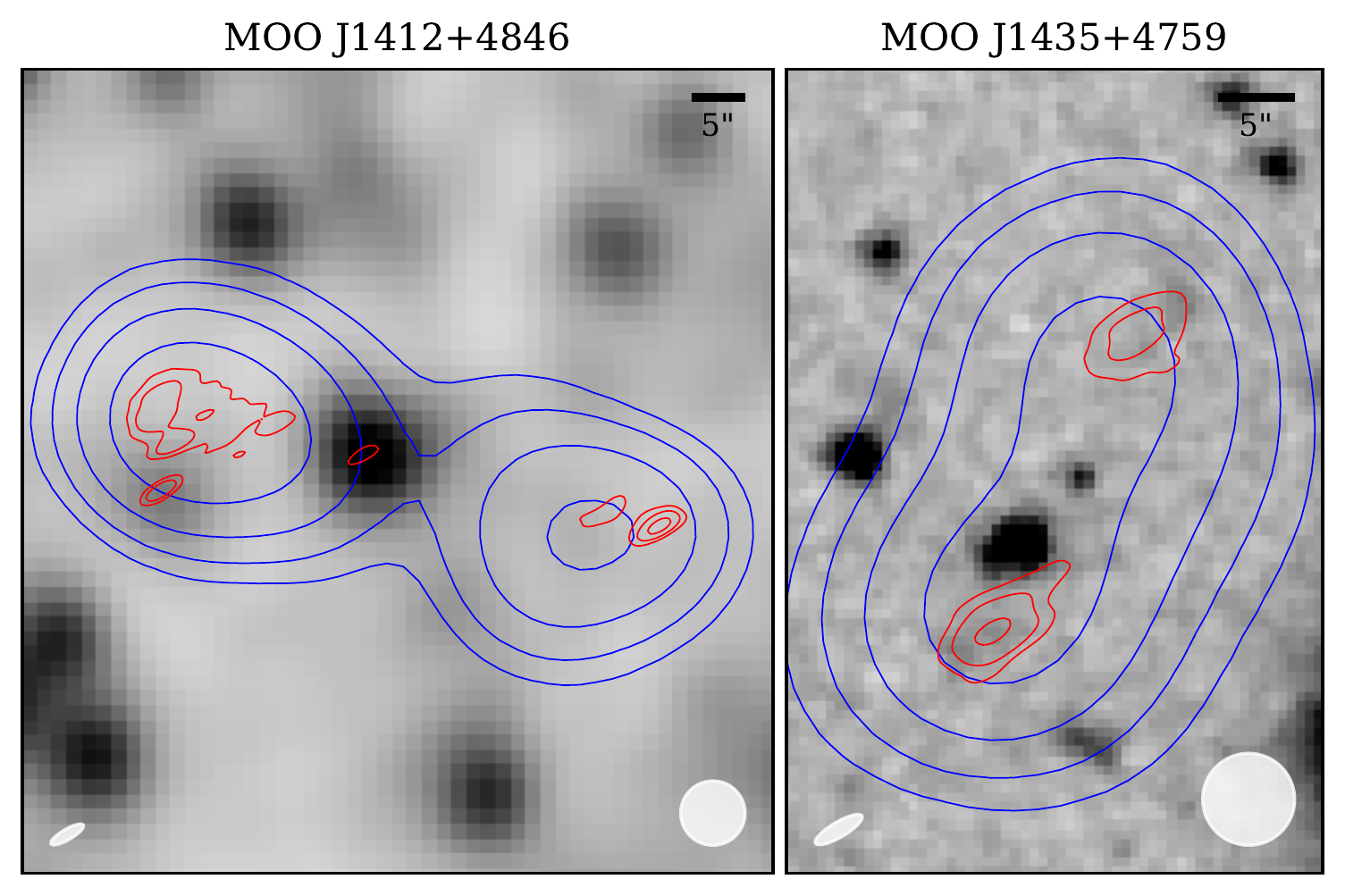}
	\caption{Two of the clusters that have LoTSS data. VLA and LoTSS contours are overlaid on a \wise\ 3.6$\mu$m image for MOO 1412+4846 and on a \spz\ 3.6$\mu$m image for MOO J1435+4759. VLA contours are shown in red starting at 4$\sigma$ and increasing by factors of 2$^n$ where n = 1,2,3, etc. LoTSS contours are shown in blue, start at 4$\sigma$, and increase in a similar fashion and are smoothed. The LoTSS synthesized beam is shown in the lower right hand corner and the VLA synthesized beam is shown in the lower left hand corner. LoTSS reveals previously unknown connected radio emission.}
	\label{fig:lotss}
\end{figure}

\begin{figure}
    \centering
	\includegraphics[width=\linewidth, keepaspectratio]{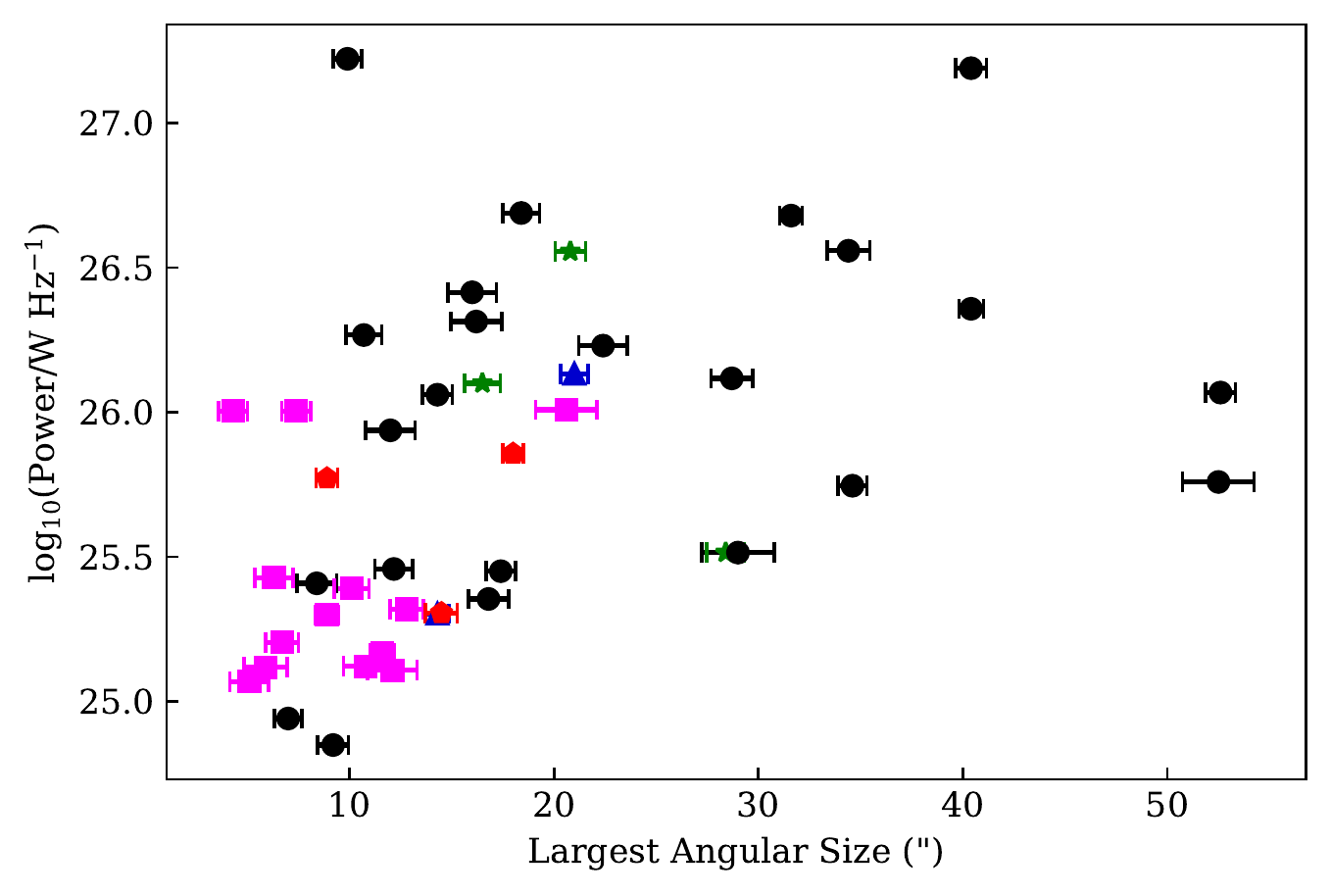}
	\caption{Radio power ($P_{1.4}$) as a function of the largest angular size (LAS). The colors and shapes are indicative of the radio morphology where a black circle is FR II, a green star is FR I, a blue triangle is FR, a red pentagon is BT, and a pink square is UD. As expected, LAS increases with luminosity.}
	\label{fig:lum_LAS}
\end{figure}

\section{Radio Morphology} \label{sect:radio}
\subsection{Classification}\label{classification}
Double-lobed radio sources are a prevalent radio morphology and are divided into two classes: Fanaroff and Riley (FR) classes I and II \citep[]{FR74}. FR I sources are `edge-darkened' in appearance in that the emission is brighter near the radio core and becomes fainter radially outward. FR II sources are `edge-brightened' in appearance in that the well-separated lobes end in distinctive areas of brightest emission (i.e. ``hotspots"). Beyond  morphological distinction, FRIs and FRIIs have other differences. Though there is still debate concerning the connection between accretion mode and jet morphology, the FR dichotomy is typically explained by a difference in jet dynamics where FRIIs have jets that remain relativistic throughout terminating in hotspots whereas FRIs are disrupted on kpc scales \citep[see][]{Mingo19,TO16}. The environment is often invoked as the cause of this disruption \citep{Kaiser07,Gendre13,Croston19}. Additionally, FR II sources are generally more luminous than FR I sources \citep[]{FR74,LO96}, though this distinction is not absolute \citep{Mingo19}. And lastly, it has been shown that the internal composition radio-lobe plasma is systematically different in these two types of radio galaxies \citep{Croston18}; the particle content of FRI lobes is dominated by protons and ions, whereas FRIIs are primarily composed of an electron-positron plasma.

With the resolution of the VLA, we were able to visually determine the radio morphologies of the sources in our sample, which are shown in Figure \ref{fig:radio_sources}. We visually classified the primary sources per the following definitions:
\begin{itemize}
    \item FR I: `Edge-darkened' meaning that the emission is brighter near the radio core and becomes fainter radially outward. Jets do not have a large amount of rounded extended emission at the ends of the jets (like that of an FR II).
    \item FR II:  `Edge-brightened' meaning that the well-separated lobes ``end" in distinctive areas of brightest emission (i.e.~hotspots). Jets end in rounded extended emission and have hotspots closer to the edge of the rounded emission than to the origin of the jet.
    \item Bent Tail (BT): An FR source (either FR I or FR II) that is bent.
    \item Undetermined (UD): Has an extended morphology that is too ambiguous to categorize and/or authors could not agree on a classification.
\end{itemize}
Four of the authors independently visually classified these sources. A majority was needed to assign a classification. In two cases the authors were split as to whether the source was an FR I or an FR II, thus we have given these sources the classification of FR I/II as the authors agree that the source morphology is one of the FR types. In three cases (MOO J0228$-$0644, MOO J0928$-$0126, and MOO J1053+1052), there was an even split between two different classifications and in these cases the first author assigned a final classification. We report these morphology classifications in Table \ref{tb:morph}.

In the sample of 50 total primary sources (from the full sample of 52 observations minus two datasets that were unusable due to data quality issues -- see \S\ref{sect:vla}), we find 26 FR IIs, 4 FR Is, 2 FR I/II s, 4 BTs, and 14 UDs.

\begin{figure*}
    \centering
	\includegraphics[width=\linewidth, keepaspectratio]{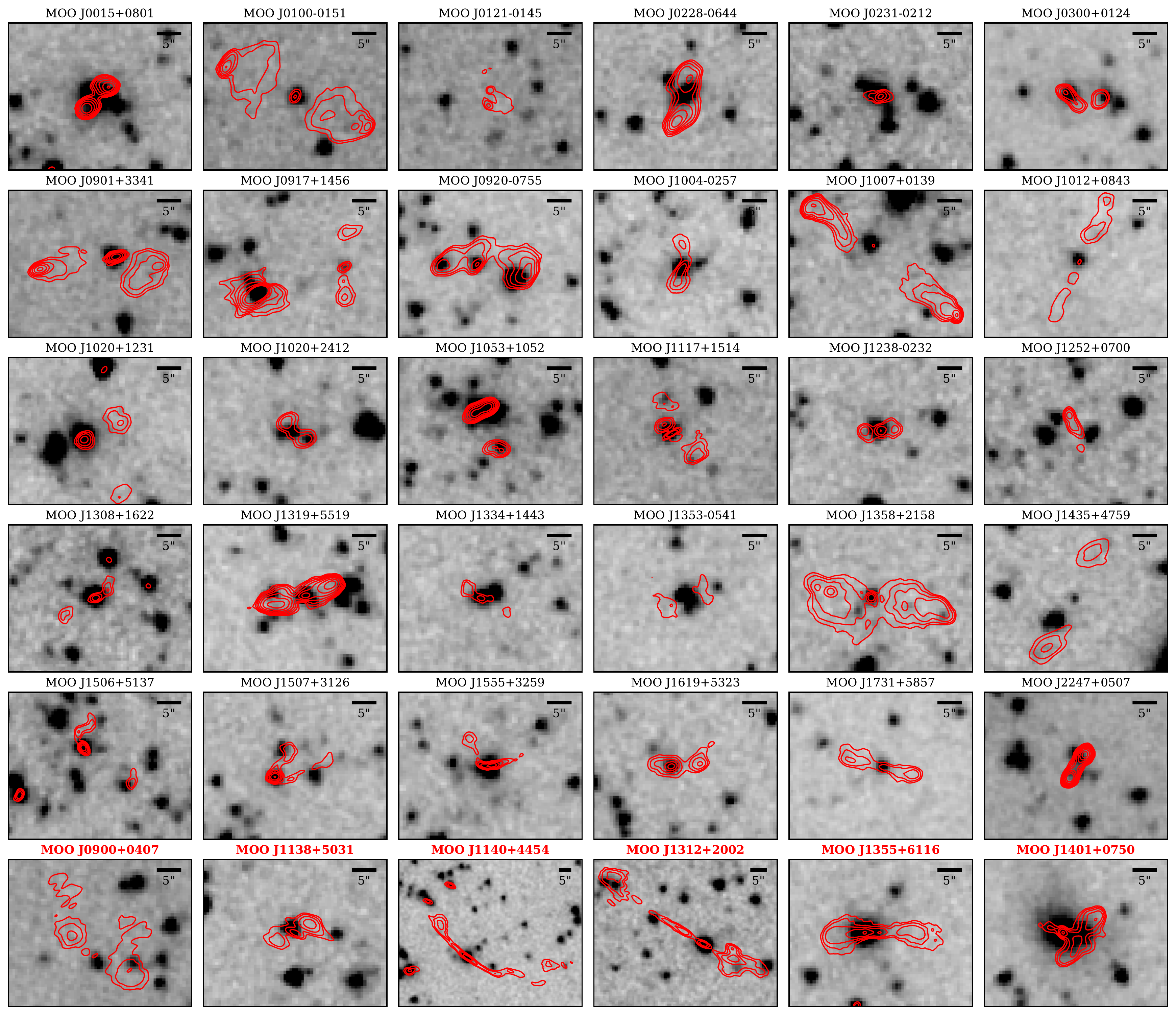}
	\caption{3.6$\mu$m images overlaid with VLA contours of the radio sources where most images are 30$^{\prime\prime}$ $\times$ 30$^{\prime\prime}$ and a 5$^{\prime\prime}$ scale bar is shown in the upper right hand corner. North is up and east is to the left. The radio contours start at 4$\sigma$ and increase by factors of 2$^n$ where n = 1,2,3, etc., except for MOO J0228$-$0644 and MOO J2247+0507 where the contours start at 8$\sigma$. For MOO J1053+1052 the source of interest is the southern source and for MOO J0917+1455 the source of interest the source to the west. Sources with red, bold titles are identified as low redshift interlopers in \S\ref{sect:interlopers}.}
	\label{fig:spz_vla}
\end{figure*}

\begin{figure}
    \centering
	\includegraphics[width=\linewidth, keepaspectratio]{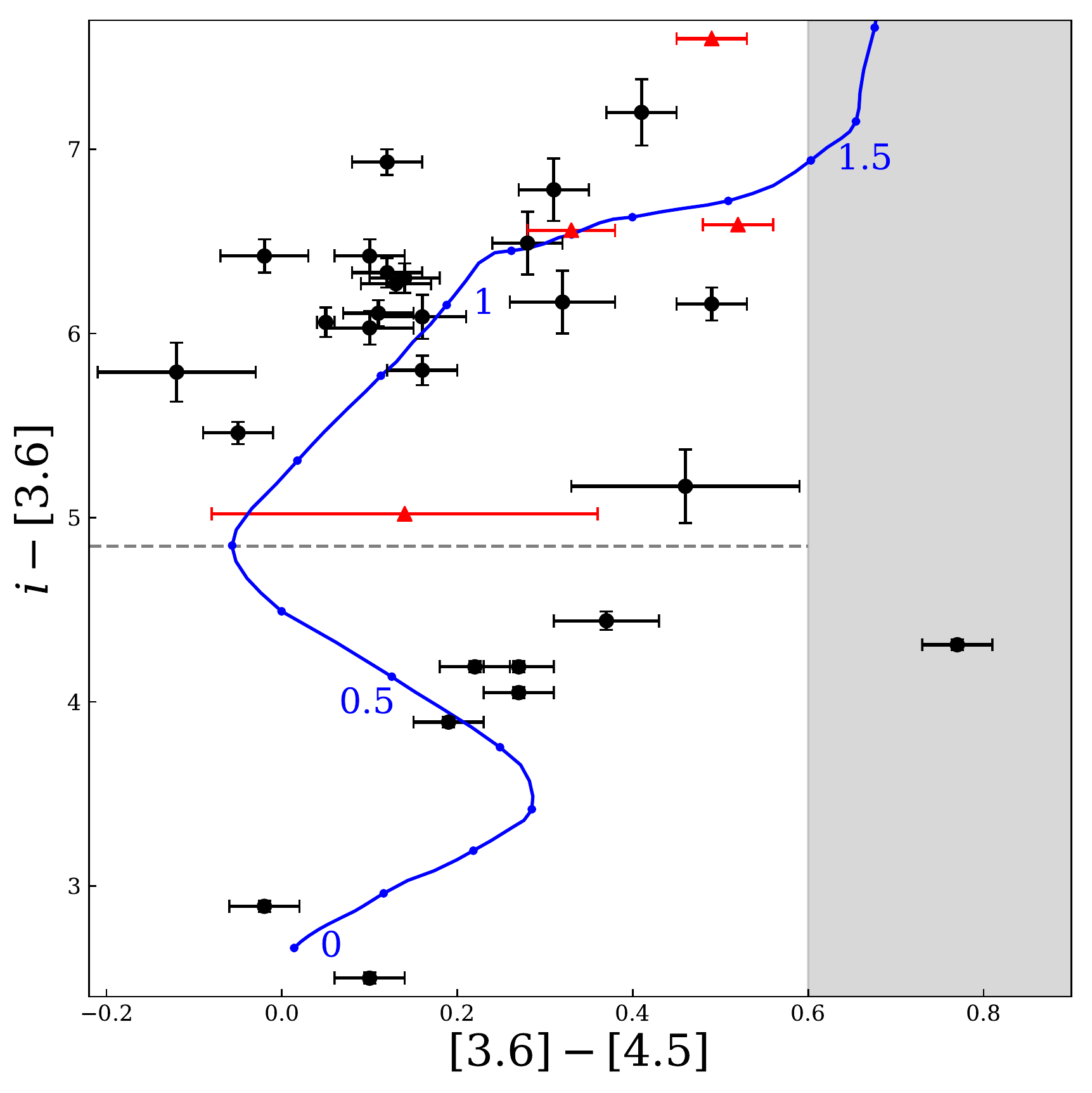}
	\caption{Color-color diagram of radio source counterparts. The blue curve indicates the expected color  evolution of a passive galaxy as a function of redshift, with each blue point marking the color at intervals of $\delta z$ = 0.1. The red triangles indicate counterparts that only have optical lower limits. The gray dashed line is the expected $i-[3.6]$ color of a passive galaxy at $z = 0.7$ and the shaded gray region represents where the color is generally dominated by emission from the AGN. The points in the lower left of the plot are flagged as low-redshift interlopers and are excluded from further analysis. The rest of the counterparts are consistent with the galaxy population expected in clusters at this redshift for this evolutionary model.}
	\label{fig:color-color}
\end{figure}

\begin{deluxetable*}{cccccc}
	\tablecaption{IRAC and Optical Counterpart Properties of the Radio Source\label{tb:cp}}
    \tablehead{\colhead{Cluster} & \colhead{[4.5]} & \colhead{[3.6] $-$ [4.5]} &\colhead{$i$} & \colhead{$i$ $-$ [3.6]} & \colhead{Stellar Mass} \\ & \colhead{} & \colhead{} & \colhead{} & \colhead{} & \colhead{($10^{11} M_{\odot}$)} }\\
	\startdata
	\multicolumn{6}{c}{\boldmath{$z > 0.7$}} \\
	\hline\\
    MOO J0015+0801 & 14.76$\pm$0.03 & 0.12$\pm$0.04 & 21.81$\pm$0.06 & 6.93$\pm$0.07 & 7.59$\pm$0.21 \\
MOO J0100$-$0151 & 15.79$\pm$0.03 & 0.49$\pm$0.04 & 22.44$\pm$0.09 & 6.16$\pm$0.09 & 2.97$\pm$0.08 \\
MOO J0228$-$0644 & 15.69$\pm$0.03 & 0.12$\pm$0.04 & 22.14$\pm$0.07 & 6.33$\pm$0.08 & 3.21$\pm$0.09 \\
MOO J0231$-$0212 & 15.72$\pm$0.03 & 0.10$\pm$0.04 & 22.24$\pm$0.08 & 6.42$\pm$0.09 & 3.67$\pm$0.10 \\
MOO J0300+0124 & 15.33$\pm$0.03 & 0.41$\pm$0.04 & 22.94$\pm$0.18 & 7.20$\pm$0.18 & 5.23$\pm$0.14 \\
MOO J0901+3341 & 14.32$\pm$0.03 & 0.77$\pm$0.04 & 19.40$\pm$0.01 & 4.31$\pm$0.03 & $-^*$ \\
MOO J0917+1456 & 17.25$\pm$0.10 & 0.46$\pm$0.13 & 22.88$\pm$0.18 & 5.17$\pm$0.2 & 0.76$\pm$0.07 \\
MOO J1004$-$0257 & 15.90$\pm$0.01 & 0.05$\pm$0.01 & 22.01$\pm$0.08 & 6.06$\pm$0.08 & 2.63$\pm$0.07 \\
MOO J1007+0139 & 16.07$\pm$0.03 & 0.16$\pm$0.04 & 22.03$\pm$0.07 & 5.80$\pm$0.08 & 2.27$\pm$0.06 \\
MOO J1012+0843 & 15.99$\pm$0.03 & 0.52$\pm$0.04 & $>$23.1 & $>$6.59 & 2.53$\pm$0.07 \\
MOO J1020+2412 & 16.27$\pm$0.04 & 0.16$\pm$0.05 & 22.52$\pm$0.12 & 6.09$\pm$0.12 & 1.86$\pm$0.07 \\
MOO J1053+1052 & 17.12$\pm$0.08 & -0.12$\pm$0.09 & 22.79$\pm$0.15 & 5.79$\pm$0.16 & 0.85$\pm$0.06 \\
MOO J1117+1514 & 15.73$\pm$0.03 & 0.31$\pm$0.04 & 22.82$\pm$0.17 & 6.78$\pm$0.17 & 3.19$\pm$0.09 \\
MOO J1238$-$0232 & 15.92$\pm$0.03 & -0.05$\pm$0.04 & 21.33$\pm$0.05 & 5.46$\pm$0.06 & 2.61$\pm$0.07 \\
MOO J1308+1622 & 15.57$\pm$0.03 & 0.14$\pm$0.04 & 22.01$\pm$0.07 & 6.3$\pm$0.08 & 3.54$\pm$0.10 \\
MOO J1319+5519 & 15.67$\pm$0.03 & 0.11$\pm$0.04 & 21.89$\pm$0.06 & 6.11$\pm$0.07 & 3.29$\pm$0.09 \\
MOO J1334+1443 & 16.21$\pm$0.04 & -0.02$\pm$0.05 & 22.61$\pm$0.08 & 6.42$\pm$0.09 & 1.98$\pm$0.07 \\
MOO J1353$-$0541 & 15.01$\pm$0.03 & 0.49$\pm$0.04 & $>$23.1 & $>$7.6 & 6.03$\pm$0.17 \\
MOO J1358+2158 & 16.35$\pm$0.04 & 0.32$\pm$0.06 & 22.84$\pm$0.17 & 6.17$\pm$0.17 & 1.73$\pm$0.06 \\
MOO J1435+4759 & 17.94$\pm$0.18 & 0.14$\pm$0.22 & $>$23.1 & $>$5.02 & 0.40$\pm$0.07 \\
MOO J1506+5137 & 16.04$\pm$0.03 & 0.28$\pm$0.04 & 22.81$\pm$0.17 & 6.49$\pm$0.17 & 2.35$\pm$0.06 \\
MOO J1507+3126 & 16.27$\pm$0.04 & 0.10$\pm$0.05 & 22.40$\pm$0.08 & 6.03$\pm$0.09 & 1.88$\pm$0.07 \\
MOO J1555+3259 & 15.24$\pm$0.03 & 0.13$\pm$0.04 & 21.64$\pm$0.04 & 6.27$\pm$0.05 & 4.83$\pm$0.13 \\
MOO J1731+5857 & 16.21$\pm$0.04 & 0.33$\pm$0.05 & $>$23.1 & $>$6.56 & 1.96$\pm$0.07 \\
\hline \\
\multicolumn{6}{c}{\boldmath{$z < 0.7$} (Likely Interlopers)} \\
\hline\\
MOO J0900+0407 & 16.34$\pm$0.04 & 0.37$\pm$0.06 & 21.15$\pm$0.030 & 4.44$\pm$0.05 & $-$ \\
MOO J1138+5031 & 15.78$\pm$0.03 & 0.10$\pm$0.04 & 18.38$\pm$0.003 & 2.50$\pm$0.03 & $-$  \\
MOO J1140+4454 & 14.83$\pm$0.03 & 0.27$\pm$0.04 & 19.29$\pm$0.005 & 4.19$\pm$0.03 & $-$  \\
MOO J1312+2002 & 14.94$\pm$0.03 & -0.02$\pm$0.04 & 17.81$\pm$0.003 & 2.89$\pm$0.03 & $-$  \\
MOO J1355+6116 & 14.43$\pm$0.03 & 0.27$\pm$0.04 & 18.75$\pm$0.004 & 4.05$\pm$0.03 & $-$  \\
MOO J1401+0750 & 14.97$\pm$0.03 & 0.19$\pm$0.04 & 19.05$\pm$0.006 & 3.89$\pm$0.03 & $-$  \\
MOO J1401+0750 & 14.11$\pm$0.03 & 0.22$\pm$0.04 & 18.52$\pm$0.004 & 4.19$\pm$0.03 & $-$  \\
\hline \\
\multicolumn{6}{c}{\textbf{No Redshift Estimates}} \\
\hline\\
MOO J0903+1319 & $-$ & $-$ & 22.75$\pm$0.22 & $-$ & $-$ \\
MOO J0928$-$0126 & $-$ & $-$ & 21.97$\pm$0.07 & $-$ & $-$ \\
MOO J0959+3903 & $-$ & $-$ & 22.05$\pm$0.04 & $-$ & $-$ \\
MOO J1011+3919 & $-$ & $-$ & 22.36$\pm$0.06 & $-$ & $-$ \\
MOO J1021+1800 & $-$ & $-$ & 21.94$\pm$0.09 & $-$ & $-$ \\
MOO J1034+3104 & $-$ & $-$ & 22.84$\pm$0.15 & $-$ & $-$ \\
MOO J1037+0433 & $-$ & $-$ & 22.32$\pm$0.08 & $-$ & $-$ \\
MOO J1412+4846 & $-$ & $-$ & 22.39$\pm$0.13 & $-$ & $-$ \\
MOO J1440+2628 & $-$ & $-$ & 21.84$\pm$0.07 & $-$ & $-$ \\
MOO J2321+0119 & $-$ & $-$ & 23.08$\pm$0.20 & $-$ & $-$ \\
	\enddata
    \tablecomments{In the first section ($z > 0.7$), we list clusters that have colors consistent with being in the cluster (see \S\ref{sect:interlopers}). In the second section ($z < 0.7$), we list clusters that have colors consistent with low redshift interlopers. In the first two sections, the Pan-STARRS \textit{i}-band counterparts are the matches of the \spz\ counterparts within 1\as\. In the last section (no IRAC), we list clusters that either do not have IRAC data or do not have an IRAC counterpart, but do have a PAN-STARRS counterpart within $\sim$2\as\ of inferred origin of the radio emission. The counterparts IRAC magnitudes are 4$^{\prime\prime}$ aperture magnitudes corrected to total magnitudes and are in Vega units. Pan-STARRS \textit{i}-band magnitudes are PSF and in AB units. The uncertainty in the stellar mass does not include systematic uncertainty related to the choice of IMF, which is about 25\%. The values in this table are based on the assumptions that the counterpart is at the photometric redshift of the cluster and that AGN emission is sub-dominant to the galaxy emission at infrared wavelengths.} 
\end{deluxetable*}

\subsection{Supplementary LoTSS Data} \label{LTS}
The LOFAR Two-metre Sky Survey (LoTSS) is an ongoing sensitive, high-resolution 120-168 MHz survey of the entire northern sky\footnote{https://www.lofar-surveys.org/releases.html}. In February 2019, the first full-quality public data release of the LOFAR Two-metre Sky Survey (LoTSS) became available, presenting  2\% of the eventual coverage \citep{Shimwell19}. The median sensitivity is $S_{\mathrm{144MHz}}$ = 71 $\mu$Jybeam$^{-1}$ and the resolution of the images is 6$^{\prime\prime}$.

Several of our targets were contained within the LoTSS DR1 (see Figure \ref{fig:lotss}). We used this low-frequency data to aid us in classifying the sources. With constant energy injection, the synchrotron radiation from an AGN produces a power law radio spectrum. However, as the AGN activity fades, energy losses cause the initial power law spectrum to steepen beyond a break frequency (whose position is related to the time since acceleration). Thus, high-frequency radio observations trace the current energy injection through the jets and hotspots, while low-frequency radio observations allow exploration of the full extent of the jet emission and the history of emission. 

In the case of MOO J1412+4846, the LoTSS data reveal extended, connected emission around the hotspots shown in the VLA data. The LoTSS data makes the FR II classification obvious, whereas it was not entirely clear from the VLA data alone. The low-frequency data also confirm the FR II classification for MOO J1435+4759. In general, the VLA data highlight the hot spots and the emission from the host galaxy and the LoTSS data allow exploration of the lobes in a different regime as the recoverable angular extent is larger at lower frequencies. Thus, we see that having lower frequency data increases the number of correctly classified FRIIs. In these cases, the VLA data, revealing only well-separated (sometimes oddly shaped) hotspots,  might lead one to classify something as undetermined, particularly if their separation is beyond the detectable largest angular size. In contrast, in the low-frequency one can clearly see the connection between the hotspots, as we see in the cases of MOO J1412+4846 and MOO J1435+4759. Overall, the LoTSS data reveals the extended emission, reducing the classification ambiguity and thus decreases the number of undetermined sources and increases the number of FRIIs, FRIs, and BTs. In order to make a one-to-one comparison, all sizes are based on the VLA data since it has higher resolution and only a few clusters have LoTSS data.

Lastly, in the case of MOO J0121$-$0145, it is possible that it might be similar to MOO J1412+4846 in that the primary and secondary sources are the hotspots of two lobes. However, there is no LoTSS data to confirm this as there is for MOO J1412+4846 (see \S\ref{LTS}). Additionally, there was no obvious optical counterpart where the host galaxy is expected to be. Thus, MOO J0121$-$0145 is classified as UD and excluded from the analysis that follows.

\subsection{Radio Extent}\label{extent}
As described in \cite{EM19}, we measure the extent of the radio sources as a diagnostic of the local environment. The largest angular size (LAS) is defined as the length of a straight line between the most distant points belonging to the same radio source. To determine the LAS, we measure the largest projected angular extent of the source contained within the lowest reliable contour. For most cases in this study the lowest reliable contour was 4$\sigma$, but for a few cases (MOO J0228$-$0644 and MOO J2247+2247), due to residual artifacts from the cleaning process, the lowest reliable contour was 8$\sigma$. These LAS measurements are given in Table \ref{tb:morph} and shown as a function of luminosity in Figure \ref{fig:lum_LAS}. We note that the measured LAS from the VLA data is a lower limit on the true LAS given that these data are high resolution and extended emission beyond $\sim$30\as\ will be resolved out. Additionally, we convert the LAS into a largest linear projected size (LLS) using a scale factor that assumes the radio source is at the cluster redshift.

\section{Host-Galaxy Properties}\label{sect:hg}
As noted in Section \ref{sect:spz}, 36 of the 50 clusters presented in this work have \textit{Spitzer} 3.6$\mu$m and 4.5$\mu$m data. We use these infrared data to investigate the radio-galaxy host-galaxy properties. We overlay VLA contours on the \spz\ images for the 36 clusters that have \spz\ data (see Figure \ref{fig:spz_vla} and Table \ref{tb:cp}) and find that 30 of the radio sources have infrared counterparts. There are several that were determined not to have counterparts. MOO J0121$-$0145 and MOO J1252+0700 obviously do not have infrared counterparts that align with the radio contours. MOO J0920$-$0755 and MOO J1619+5323 do not have counterparts that line up with inferred origin of the radio emission where one would expect the counterpart for radio galaxies to be. The source of interest in MOO J1020+1231 is the source in the upper right and this one does not have an infrared counterpart. MOO J2247+0507 has two infrared counterparts that are encompassed by the radio contours, however neither are unambiguously associated with the inferred origin of the radio jets. Thus, none of the sources detailed above are listed as having counterparts in Table \ref{tb:cp}.

\begin{figure*}
    \centering
	\includegraphics[width=\linewidth,keepaspectratio]{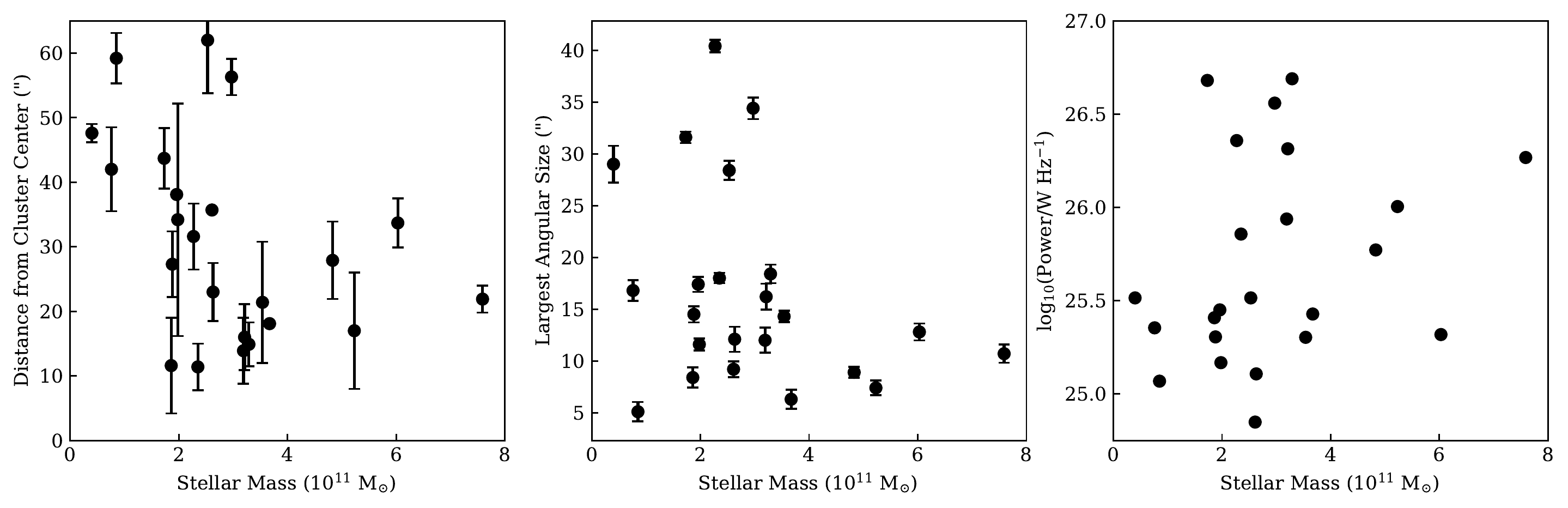}
	\caption{Correlations between the stellar mass of the host galaxy, distance from the cluster center, LAS, and power. Only the host galaxies with colors consistent with being in the cluster are shown. There is an inherent $\sim$25\% error in the stellar mass due to the choice of IMF (see \S\ref{sect:SM}) and is not shown. In combination, these correlations indicate that the compact, highest stellar mass galaxies are confined to the center of the cluster.}
	\label{fig:SM}
\end{figure*}

\subsection{Interloper Identification} \label{sect:interlopers}
We can identify low redshift interloper sources by comparing the observed infrared colors to those of a model passively evolving galaxy. For those sources that have infrared counterparts, we use a 1\as\ matching radius to obtain the \textit{i}-band photometry from Pan-STARRS and calculate the observed-frame $[3.6] - [4.5]$ and $i-[3.6]$ colors (see Table \ref{tb:cp}). In the cases of non-detections in Pan-STARRS, we assume that the counterpart has an \textit{i}-band magnitude fainter than the limiting magnitude \textit{i} = 23.1 \citep[AB;][]{PS16} and place a lower limit on the \textit{i}$-$[3.6] color. We use the PS1 DR2 stacked PSF magnitudes and positions.\footnote{We note that in \cite{EM19}, the Pan-STARRS magnitudes had an incorrect conversion applied making them brighter than they should be by $\sim$0.7 magnitudes. However, this error did not change key results in that paper, and is corrected in this work.} In the case that there are two magnitudes at one position due to overlaps in the stack tessellations, we adopt the one with labeled as the primary detection. 

We compare the observed colors to the expected colors of a passively evolving galaxy as a function of redshift using \verb'EzGal' \citep[]{EzGal12}. Following the parameters used in \cite{Gonzalez19} for \madcows, we use an FSPS model \citep[]{c09} with a simple stellar population, a \cite{Chab03} initial mass function (IMF), and a formation redshift of $z_f$ = 3. Because the brightest galaxies in clusters have super-solar metallicities \citep{Connor19}, we also assume $Z=0.03$.

In Figure \ref{fig:color-color}, we plot the counterparts compared to the evolutionary track to identify the counterparts that are low redshift interlopers (the dashed gray line is the color expected for a passive galaxy at $z$ = 0.7) or those whose emission is dominated by an AGN (shown by the shaded region; \citealt{Stern12}). When comparing the colors of the host galaxies to the evolutionary track, we find that seven counterparts (two are associated with MOO J1401+0750) have evidence of being low redshift interlopers and we exclude these from the following analysis (marked as likely interlopers in Table \ref{tb:cp}). We find one counterpart (MOO J0901+3341) whose color is indicative of being AGN-dominated rather than stellar-dominated (\citealt{Stern12}). Thus, we do not calculate the stellar mass for this object. We do, however, include it in analysis that does not pertain to stellar mass. In total, we identify 23 as being consistent with being in the cluster, 6 low redshift interlopers, and one that has no redshift estimate due to being dominated by AGN emission. We note that spectroscopy will be required to better estimate the true contamination level.

\subsection{Stellar Masses and Candidate BCGs}\label{sect:SM}
Under the assumptions that (1) these counterparts are cluster members and (2) the emission of the AGN is sub-dominant to that of the galaxy at these mid-infrared wavelengths, we can use IRAC photometry to estimate stellar masses (supported by the color analysis above). We convert [4.5] to a stellar mass using \verb'EzGal' assuming a Flexible Stellar Population Synthesis (FSPS) model \citep[]{c09} with a simple stellar population, a formation redshift $z_f$ = 3, a super-solar metalliticity $Z=0.03$, a \cite{Chab03} initial mass function (IMF), and the cluster photometric redshift. The stellar mass changes by $\pm$ 25\% for $2\leq z_f \leq 5$. We report the stellar masses in Table \ref{tb:cp} for a formation redshift $z_f$ = 3.

The average stellar mass is 2.9$\times$10$^{11}$ M$_{\odot}$ with 21 out of the 23 galaxies having $M_*$ $\geq$ 10$^{11}$ M$_{\odot}$. The highest mass is 7.6$\times$10$^{11}$ M$_{\odot}$ and the lowest is 0.4$\times$10$^{11}$ M$_{\odot}$.

Using the \spz\ data, we can determine candidate brightest cluster galaxies (BCGs) within each cluster. We identify the candidate BCG using the following procedure. First, we identify the galaxies that are not dominated by AGN emission \citep{Stern12} by applying a color cut of [3.6] $-$ [4.5] $<$ 0.6. Then, for the 30 brightest in [4.5] we find Pan-STARRS cross matches within 1\as\ using \verb'astroquery' \citep{Ginsburg19}. We label the galaxy with the brightest [4.5] magnitude and $i$ $-$ [3.6] greater than the color expected for a passive galaxy at $z$ = 0.7 (4.85; calculated using \verb'EzGal') the brightest cluster galaxy. Out of the 23 counterparts consistent with the galaxy population expected in clusters at $z>0.7$, nine of them are the candidate BCG. In total, 12 of the candidate BCGs have radio emission associated with them. One of the BCGs (MOO J0901+3341) has a [3.6] $-$ [4.5] that is consistent with being AGN dominated which qualifies it as an optically bright quasar. As such, though its [4.5] is consistent with being the BCG, we exclude it from this BCG analysis.
%Out of the 23 counterparts consistent with the galaxy population expected in clusters at $z>0.7$, four of them are the candidate brightest cluster galaxy. 

A total of 19 out of the 23 counterparts (83\%) are consistent with being one of the six most massive galaxies in the cluster. Predictably, all of the five most massive host galaxies in this sample are the candidate brightest galaxy in the cluster. These observations are consistent with the expectation that extended radio lobes are predominantly hosted by massive galaxies \citep[e.g.,][]{Seymour07}. They are also consistent with previous studies that find that the prevalence and strength of radio activity is a strong function of stellar mass, hence BCGs are likely to be radio active \citep{Best05,Janssen12,Sabater19}. BCGs would therefore be expected to be more likely to be radio-active based upon stellar mass alone.
%Though it has also been found that BCGs are more likely to be radio-loud compared to other cluster galaxies at a similar stellar mass \citep{Best07}. }

\subsection{Trends with Stellar Mass}
We investigate correlations between stellar mass and both the distance of the source from the cluster center and LAS (see Figure \ref{fig:SM}). We observe $M_{*} <$ 4$\times 10^{11} M_{\odot}$ radio hosts at all radii, but $M_{*} >$ 4$\times 10^{11} M_{\odot}$ only at distances less than 40$^{\prime\prime}$ ($\sim$325 kpc) from the cluster center, indicating that the highest stellar mass hosts are confined to the cluster center. This could be explained by mass segregation in which more massive galaxies are preferentially found near the cluster center \citep[e.g.,][]{Abell67,Quintana79,Roberts15} or that if the host of the radio galaxy is a BCG it will be preferentially near the center of the cluster. 

We find a similar trend with respect to largest angular size. We observe $M_{*} <$ 4$\times 10^{11} M_{\odot}$ radio hosts at all sizes, but $M_{*} >$ 4$\times 10^{11} M_{\odot}$ only at sizes less than 15$^{\prime\prime}$ ($\sim$120 kpc). Taken together, these correlations with stellar mass indicate that the highest stellar mass galaxies host compact jets and tend to be in the inner portion of the cluster. 

It is well-known that the ICM found in galaxy clusters has a decreasing density profile radially outward \citep{Cavaliere78,Vikhlinin06,Croston08}. It is also known that the jet length of AGN depends on the density of the medium that it is in, the power of the AGN, and its age \citep{Falle91,Alexander00,Kaiser07}. The result that the highest stellar mass galaxies host compact jets and tend to be in the inner portion of the cluster can be understood in the context of this decreasing ICM density profile found in galaxy clusters \citep{EM19}. It is known that higher density ICM like that found in the center of a cluster will more efficiently pressure confine jets than lower density ICM like that found at the outskirts of a cluster \citep{Kaiser97,Alexander00,Kaiser07}. Thus, one explanation for the smaller jets of the higher stellar mass galaxies at the center of the cluster is that these jets are being more confined by the denser ICM compared to jets farther from the cluster center \citep{EM19}.

This interpretation is strengthened by the fact that radio power does not increase with stellar mass (Spearman's rank-order correlation $r_s$ = 0.23, $p$ = 0.29), which has been noted previously by \cite{Best07}, \cite{L&M07}, and \cite{Gupta19} (see Figure \ref{fig:SM}). As noted in \S\ref{sect:LAS_v_D}, the size of the jets depends on power, the density of the surrounding medium, and the age of the source. This lack of correlation between stellar mass and power indicates that the power does not affect the relationship between stellar mass and angular size, leading to the conclusion that there are perhaps environmental factors affecting the size of these sources, as described above.

\section{Radio Properties and Environmental Factors}\label{sect:environ}
With the radio morphologies defined and interlopers identified, we can investigate whether the properties of the radio emission (e.g., luminosity and LAS) correlate with properties of the environment (e.g., richness and distance from the cluster center). The (non-)existence of correlations between these properties is a diagnostic of the astrophysics at work within these environments. If the environment affects the evolution of the radio galaxy, one would expect to see a correlation between these properties \citep[e.g.][]{Best07,Ineson15,Croston19}. In \S\ref{sect:LAS_v_D}, we revisit the correlation between LAS and distance from the cluster center seen in \cite{EM19}. In \S\ref{sect:lum}, we investigate the relationship between the luminosity and the distance from the cluster center. In \S\ref{sect:rich_lum}, we investigate trends with other environmental properties (distance from the cluster center and richness). We perform Spearman's rank-order correlation on the relationships, producing the rank order coefficient ($r_s$) and the p-value ($p$).

\begin{figure}
    \centering
	\includegraphics[width=\linewidth,keepaspectratio]{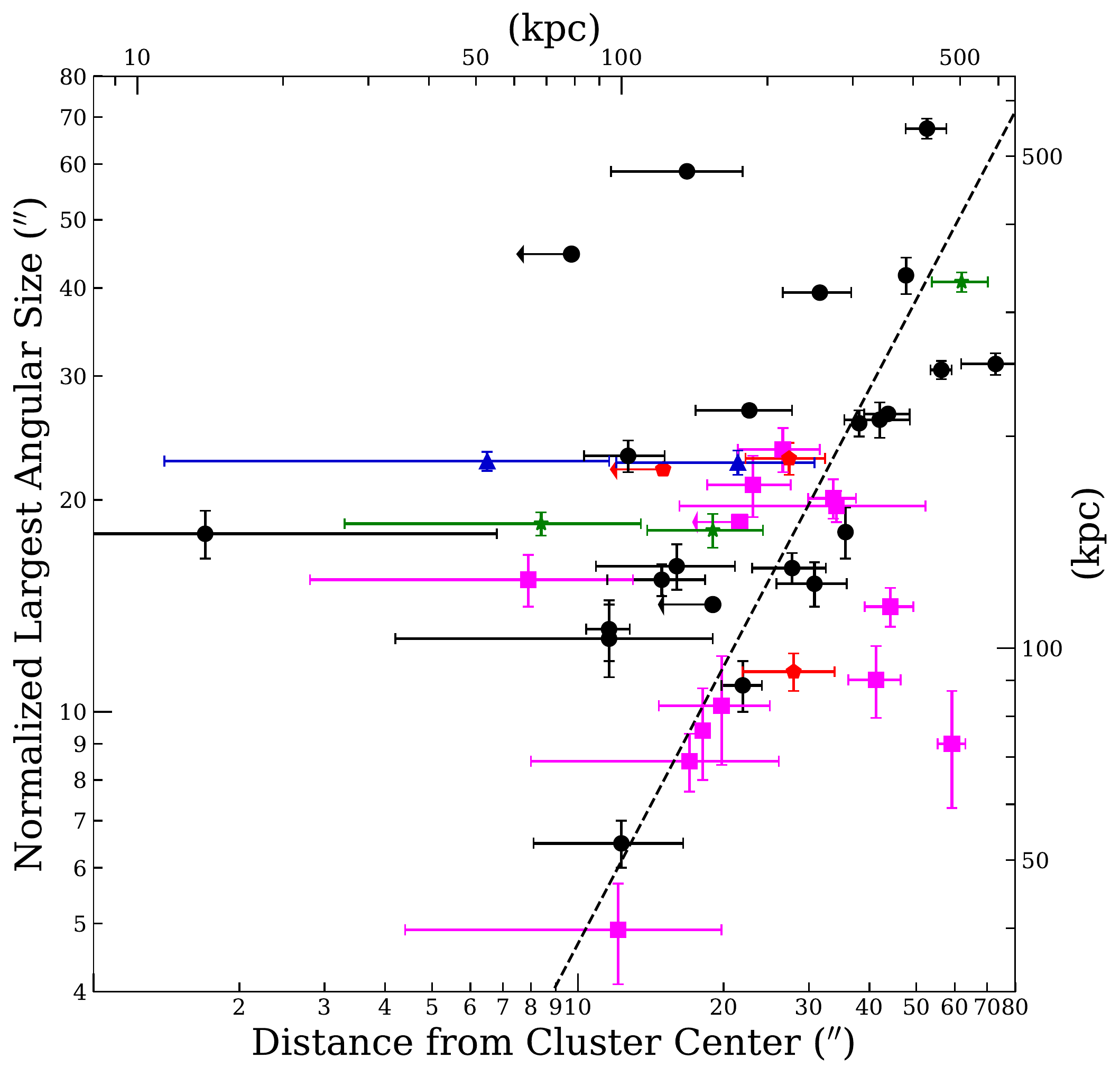}
	\caption{The relationship of largest angular size (arcseconds) versus the average distance from the cluster center for the full sample. The distance is the average distance in arcseconds of the radio source to the cluster center using the \wise\ and \textit{Spitzer} centers. The top axis is the distance from the cluster in arcseconds converted to kpc assuming $z=1$. The colors and shapes are indicative of the radio morphology where a black circle is FR II, a green star is FR I, a blue triangle is FR, a red pentagon is BT, and a pink square is UD. Arrows represent the sources that, with error bars, are consistent with being at the cluster center. The dashed line is the relationship from \cite{EM19}.}
	\label{fig:LAS_pos}
\end{figure}

\subsection{Radio Extent and Distance from the Cluster Center}\label{sect:LAS_v_D}
In \cite{EM19}, we observe a strong correlation between the largest angular extent and the distance from the cluster center. In this work, we are able explore this relation in depth given the larger sample size. After applying the color cut described in \S\ref{sect:interlopers}, there are 44 sources available to investigate this relation (see Figure \ref{fig:LAS_pos}). As in \cite{EM19}, we normalize the size to a fiducial power according to 
\begin{equation}\label{eqn:LAS}
    L = c \bigg(\frac{P}{\rho_0(R)}\bigg)^{1/5}t^{3/5} 
\end{equation}
such that 
\begin{equation}
    L_{\mathrm{norm}} = L \bigg(\frac{P_0}{P_{\mathrm{1.4}}}\bigg)^{1/5}
\end{equation}
where $P_0$ = 2$\times$10$^{26}$ W Hz$^{-1}$ using the $P_{1.4}$ from Table \ref{tb:morph}. 

The distance of the source from the cluster center is defined as the angular distance between the inferred origin of the radio emission based on the morphological classification and the cluster center. The distance of the radio source from the center of the cluster for all morphology types, plotted in Figure \ref{fig:LAS_pos}, is the average of the distance of the source from the {\it WISE} and {\it Spitzer} cluster centers (see \S\ref{sect:spz}). We take the uncertainty in the distance from the cluster center to be the difference between the two distances and dividing by two. For those clusters that do not have a robust \textit{Spitzer} center, we adopt the \wise\ center as the center and the error in the distance to be an average of the errors of the points that had robust \spz\ centers.

The scatter of this relationship visually increases from that in \cite{EM19} when the full sample of radio sources consistent with being high redshift is considered. While there exists an overdensity of sources along the previous relation, a substantial subset now lie to the upper left, off the relation. Projection of the radio galaxy locations could explain these points (which we explore below). Likewise, the points to the lower right of the relation could be explained by projection of the jets to lower apparent angular sizes. This subset could also be explained by unidentified low redshift interlopers (not all clusters had \spz\ data). Interlopers would contaminate the plot by adding sources that have a larger LAS than predicted by the relation because sources are inherently larger at a lower redshift. For completeness, we note that projection could also shift a source onto the relation.

We ran simulations to investigate whether projection effects can explain the distribution of the data. Below are the steps of the projection simulations:
\begin{enumerate}
    \item Assume that the pilot relation still holds and randomly choose a distance ($d$) on the relation.
    \item Randomly choose a projected point on a unit sphere and scale it by $d$ to obtain a projected distance, $d_{\mathrm{obs}}$.
    \item Scatter $d_{\mathrm{obs}}$ using a Gaussian distribution with a standard deviation of 5.1$^{\prime\prime}$, the mean error of the observed cluster center.
    \item Find the corresponding LAS for the original unprojected distance.
    \item Project the LAS to account for the orientation using a similar approach to above, but scatter it by  1.2$^{\prime\prime}$, the mean error of largest angular sizes.
    \item For any simulated point that has a LAS less than 6.5$^{\prime\prime}$, regenerate the point as the size threshold imposed in this study is greater than 6.5$^{\prime\prime}$ (see \S\ref{sample}).
\end{enumerate}

\begin{figure}
    \centering
	\includegraphics[width=\linewidth,keepaspectratio]{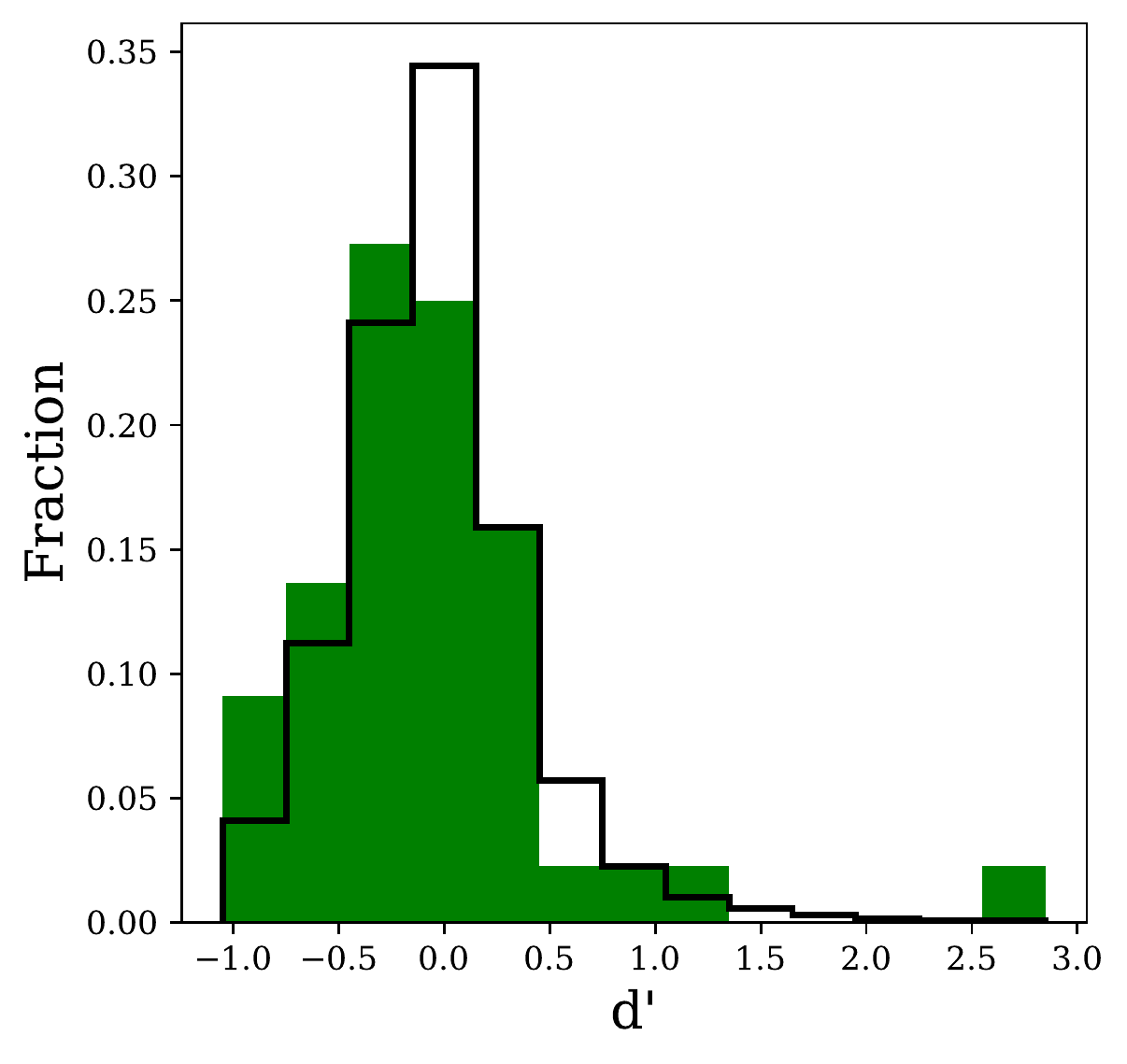}
	\caption{Green is the $d^{\prime}$ distribution of the data and black is the distribution of 100,000 randomly projected points described in \S\ref{sect:LAS_v_D} where $d^{\prime}$ is the normalized distance of the points from the \cite{EM19} relation. Due to the similarity of these distributions, we find that projection accounts for a majority of the spread in the points of Fig \ref{fig:LAS_pos}, however the excess at $d^{\prime}$ $\lesssim$ -0.2 could be a distinct population (see \S\ref{sect:LAS_v_D}).}
	\label{fig:dsim}
\end{figure}

To compare these simulations to the data, we define a normalized distance 
\begin{equation}
    d^{\prime} = \frac{d_{\mathrm{obs}}}{d_{\mathrm{relation}}} - 1 
\end{equation}
% \begin{equation}
%     d^{\prime} = \frac{\begingroup\color{blue}d_{\mathrm{obs}}\endgroup-\begingroup\color{red}d_{\mathrm{relation}}\endgroup}{\begingroup\color{red}d_{\mathrm{relation}}\endgroup}
% \end{equation}
where $d_{\mathrm{obs}}$ is the distance from the cluster center of the particular data point and $d_{\mathrm{relation}}$ is the corollary distance of $d_{\mathrm{obs}}$ on the pilot relation. In Figure \ref{fig:dsim}, we compare the normalized distribution of $d^{\prime}$ values for the data (green histogram) to the normalized distribution of the 100,000 simulated projected points (black histogram outline).

Compared to the simulation, the observed distribution of radio sources is similar in shape (see Figure \ref{fig:dsim}) around $d^{\prime}$ = 0. To quantify the similarity of the two distributions, we perform a two sample Kolmogorov-Smirnov test on the simulated data and the real data. We obtain $p$=0.086, which is above the typical cutoff value of 0.05, above which the data are consistent with the null hypothesis. We choose a bin size of 0.3 in order to minimize Poisson noise in the most populated bins and showcase the similarity between the distributions. The distribution of the data is undoubtedly influenced by unidentified low redshift interlopers. For instance, the removal of the most extreme outlier at low distance from the cluster center increases the p-value to 0.014. Thus, we consider the two distributions to be consistent with originating from the same underlying distribution. 

Compared to the simulation, there is an excess of sources on the order of a few percent at $d^{\prime} \lesssim -0.2$. This excess of sources is a population of radio-AGN that have a large angular extent, but are near the cluster center. One possible explanation is that there are interlopers at lower redshift. If not, a physical explanation might be that these are cases in which AGN activity at the center of the cluster creates bubbles of low pressure (\citealt{Blanton10}). Over time the jets continue to inject energy episodically into the ICM, the bubbles of low pressure expand, and thus the jets are able to expand further into the ICM and have a larger extent. 

Overall, we conclude that the effects of projection account for a majority of the spread in angular size that is seen in this work. A future improvement to increase the certainty in the origin of the scatter would be to obtain spectroscopic redshifts for these sources to identify and discard any interlopers.

\begin{figure*}
    \centering
	\includegraphics[width=\linewidth,keepaspectratio]{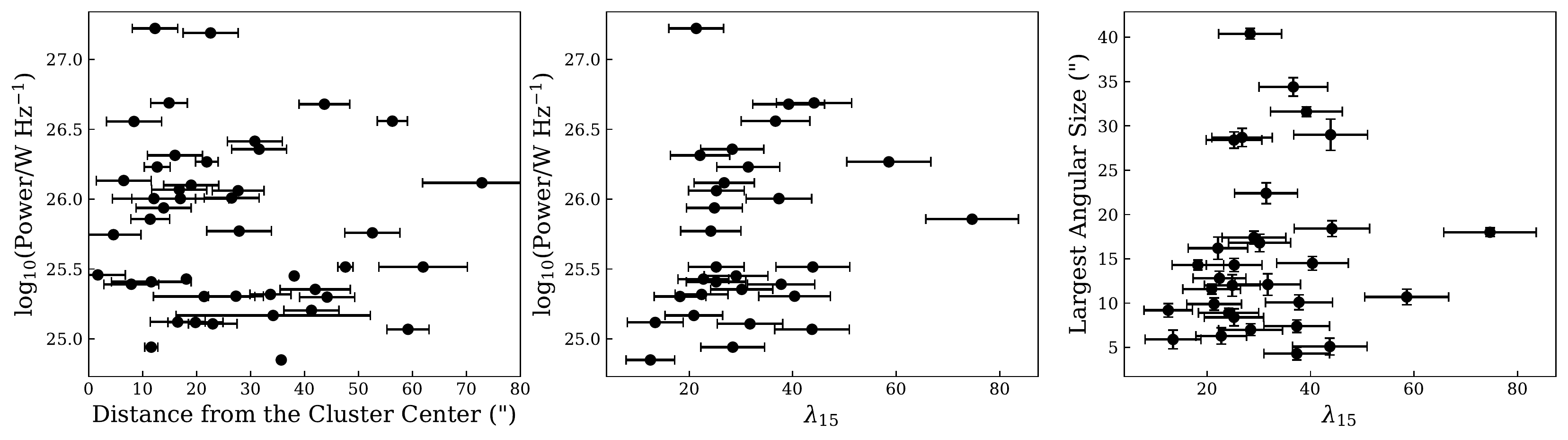}
	\caption{Correlations between the radio-AGN properties (luminosity and LAS) and properties of the environment such as richness ($\lambda_{15}$) and distance from the cluster center. We find no correlations between these properties (see \S\ref{sect:environ}).}
	\label{fig:environ}
\end{figure*}

\subsection{Luminosity and Distance from the Cluster Center}\label{sect:lum}
We investigate the correlation between luminosity and distance of the source from the cluster center (see Figure \ref{fig:environ}) and see no evidence of a correlation ($r_s$ =$-$0.15, $p$ = 0.35). In contrast, in a study of radio galaxies associated with clusters at $z<0.4$ using LoTSS, \cite{Croston19} find that luminosity increases towards the cluster center. One explanation for this discrepancy in trends at different redshift ranges is that radio galaxies at high-redshift could be triggered stochastically based on the availability of fuel in a more dynamically evolving environment (this work), while at lower redshift the galaxies have settled into a more evolved state where the AGN activity is more linked to the environment (\citealt{Croston19}).

\subsection{Relationship with Environmental Properties}\label{sect:rich_lum}
We investigate whether there are any relationships between radio source properties and the richness of the cluster ($\lambda_{15}$) in Figure \ref{fig:environ}. There is a tentative suggestion of a trend of increasing luminosity with richness, but this correlation is visually weak. We quantify the weakness of this correlation by calculating Spearman's rank coefficient ($r_s$ = 0.27) and the p-value ($p$ = 0.15) which show that the data are consistent with the null hypothesis that there is no association between the two variables. 

In the literature, varying degrees of strength of a correlation between luminosity and cluster properties have been seen at a range of redshifts. At $z<0.5$, there is strong evidence for a correlation between luminosity and measures of the density of the environment which is measured by galaxy counts and X-ray luminosity \citep{Best04,Ineson13,Ineson15,Ching17,Croston19}. However, at higher redshift ($ 0.5 \lesssim z \lesssim 1.0 $), the existence of a relationship between these two quantities is still unclear and studies have found evidence for and against correlation \citep{Wold00,Belsole07,Falder10,Wylezalek13}. Overall, there is clear evidence of correlations between power and the cluster environment at $z \lesssim 0.5$, but at $z>0.5$ a clear, uncontested relationship has not emerged. These studies and our results could be congruent with the idea posed in \S\ref{sect:lum} that radio galaxies at high-redshift could be triggered stochastically based on the availability of fuel in a more dynamically evolving environment. At lower redshift the galaxies would then have settled into a more evolved state where the AGN activity is more linked to the environment.

We find no correlation between richness and LAS ($r_s$ = 0.26, $p$ = 0.15), which is similar to the results of low redshift studies \citep[][]{Croston19}. This could be expected given the number of factors that contribute scatter to the LAS, which will dilute and weaken any relationship between richness and LAS. We note that this analysis does not consider the entire radio source population, only those that are extended which limits the range of LAS probed in this work.

\section{Summary}\label{summary}
In this work, we investigate the radio morphologies of extended radio sources and the properties of their host galaxies in 50 massive galaxy clusters at $z \sim 1$. These clusters are drawn from a parent sample of \textit{WISE}-selected galaxy clusters that were cross-matched with the VLA FIRST survey to identify extended radio sources within 1$^{\prime}$ of the cluster centers. Using radio (1.4 GHz and 6.0 GHz from the VLA), infrared (3.6$\mu$m and 4.5$\mu$m from \textit{Spizter}), and optical ($i$-band from PAN-STARRS) data, we determine the radio morphologies and investigate the role of environmental factors on the radio-loud AGN population. Below is a summary of our findings:
\begin{itemize}
    \item{\textit{Radio Morphologies:} In the sample of 50 total primary sources (from the full sample of 52 observations minus two datasets that were unusable due to data quality issues), we find 26 FR IIs, 4 FR Is, 2 FRI/IIs, 4 bent tails, and 14 undetermined radio morphologies.}
    \item{\textit{Host Galaxies:} 36 out of the 52 clusters have \spz\ data. We use the optical and infrared colors of the counterparts to identify 6 counterparts that are low redshift interlopers or whose \spz\ fluxes are dominated by emission from AGN. Of the remaining counterparts detected in \spz\, we find that the average stellar mass is 2.9$\times$ 10$^{11}$M$_{\odot}$, which is consistent with the expectation that extended radio lobes are predominantly hosted by massive galaxies \citep[e.g.,][]{Seymour07}. We also find that $\sim$40\% of the host galaxies are the candidate BCG and $\sim$83\% are consistent with being one of the top six most massive galaxies in the cluster.}
    \item{\textit{Correlation between radio source size and distance from the cluster center:} We see evidence for an increase in the size of the jets with cluster-centric radius, which can be understood as arising due to the decreased ICM pressure confinement with increasing radius. Projection effects explain a majority of the scatter of this relation.}
    \item{\textit{Correlations between environment and radio galaxy characteristics:} We find that the highest stellar mass galaxies are compact and tend to be in the inner portion of the cluster. We find no significant correlations between the AGN power or size and environmental properties such as cluster richness and location within the cluster.}
\end{itemize}
 
Even though there is evidence for a significant overdensity of AGN and radio sources within clusters (\madcows, \citealt{Mo18}; LoTSS, \citealt{Croston19}), this work finds that there is no preference for particular cluster or host-galaxy properties at this epoch. Mo et al. (in prep.) find that the fraction of clusters that contain a radio source and the fraction of cluster galaxies that are radio-active increase with cluster richness. These findings, combined with the lack of correlations with cluster properties found in this work, indicate that the cluster environment fosters radio activity, especially for the massive galaxies near the center, but does not solely drive the evolution of these sources at this redshift. To the extent that the large scale environment affects the evolution of these sources, the driver is the physical distance of the source from the cluster center showcased by the fact that the most massive radio-galaxy host galaxies are near the cluster center and host compact jets. However, the rest of the radio source population does not show a strong dependence on environmental factors. Overall, the cluster environment fosters radio-activity, but the small scale environment is driving the physics of fueling the jet and the source evolution.

\acknowledgements
E.M. would like to thank those at the NRAO workshops and helpdesk for invaluable help with imaging. 
%E.M. would like to thank the anonymous referee for their helpful and insightful comments.

This research was supported in part by the National Science Foundation (AST-1715181). This publication was made possible through the support of NASA  through the Florida Space Grant Consortium via the NASA Florida Space Grant Consortium Doctoral Improvement Fellowship. This research was also supported in part by NASA through the NASA Astrophysical Data Analysis Program, award NNX12AE15G. Parts of this work have also been supported through NASA grants associated with the \textit{Spitzer} observations (PID 90177 and PID 11080). Basic research in radio astronomy at the U.S. Naval Research Laboratory is supported by 6.1 Basic Research. 

This publication makes use of Common Astronomy Software Applications (CASA) package \citep[]{casa}.~This publication makes use of Astropy, a community-developed core Python package for Astronomy \citep[]{astropy} and APLpy, an open-host plotting package for Python \citep[]{aplpy}.

This publication makes use of data products from NSF's Karl G. Jansky Very Large Array (VLA). The National Radio Astronomy Observatory is a facility of the National Science Foundation operated under cooperative agreement by Associated Universities, Inc. Additionally, this publication makes use of data products from the \textit{Wide-field Infrared Survey Explorer}, a joint project of the University of California, Los Angeles, and the Jet Propulsion Laboratory/California Institute of Technology, funded by NASA. This work is also based in part on observations made with the \textit{Spitzer Space Telescope}, which is operated by the Jet Propulsion Laboratory, California Institute of Technology under a contract with NASA. 

%%%%%%%%%%%%%%%%% APPENDICES %%%%%%%%%%%%%%%%%%%%%
%\appendix

%%%%%%%%%%%%%%%%%%%%%%%%%%%%%%%%%%%%%%%%%%%%%%%%%%
%%%%%%%%%%%%%%%%%%%%%%%%%%%%%%%%%%%%%
\bibliography{Moravec19_RadioActiveCows_arXiv}{}

\begin{thebibliography}{}
\expandafter\ifx\csname natexlab\endcsname\relax\def\natexlab#1{#1}\fi

\bibitem[{{Abell} \& {Richard}(1967)}]{Abell67}
{Abell}, G.~O., \& {Richard}, R. 1967, \aj, 72, 783

\bibitem[{{Alberts} {et~al.}(2016){Alberts}, {Pope}, {Brodwin}, {Chung},
  {Cybulski}, {Dey}, {Eisenhardt}, {Galametz}, {Gonzalez}, {Jannuzi},
  {Stanford}, {Snyder}, {Stern}, \& {Zeimann}}]{Alberts16}
{Alberts}, S., {Pope}, A., {Brodwin}, M., {et~al.} 2016, \apj, 825, 72

\bibitem[{{Alexander}(2000)}]{Alexander00}
{Alexander}, P. 2000, \mnras, 319, 8

\bibitem[{{Astropy Collaboration} {et~al.}(2013){Astropy Collaboration},
  {Robitaille}, {Tollerud}, {Greenfield}, {Droettboom}, {Bray}, {Aldcroft},
  {Davis}, {Ginsburg}, {Price-Whelan}, {Kerzendorf}, {Conley}, {Crighton},
  {Barbary}, {Muna}, {Ferguson}, {Grollier}, {Parikh}, {Nair}, {Unther},
  {Deil}, {Woillez}, {Conseil}, {Kramer}, {Turner}, {Singer}, {Fox}, {Weaver},
  {Zabalza}, {Edwards}, {Azalee Bostroem}, {Burke}, {Casey}, {Crawford},
  {Dencheva}, {Ely}, {Jenness}, {Labrie}, {Lim}, {Pierfederici}, {Pontzen},
  {Ptak}, {Refsdal}, {Servillat}, \& {Streicher}}]{astropy}
{Astropy Collaboration}, {Robitaille}, T.~P., {Tollerud}, E.~J., {et~al.} 2013,
  \aap, 558, A33

\bibitem[{{Becker} {et~al.}(1994){Becker}, {White}, \& {Helfand}}]{Becker94}
{Becker}, R.~H., {White}, R.~L., \& {Helfand}, D.~J. 1994, in Astronomical
  Society of the Pacific Conference Series, Vol.~61, Astronomical Data Analysis
  Software and Systems III, ed. D.~R. {Crabtree}, R.~J. {Hanisch}, \&
  J.~{Barnes}, 165

\bibitem[{{Becker} {et~al.}(1995){Becker}, {White}, \& {Helfand}}]{Becker95}
{Becker}, R.~H., {White}, R.~L., \& {Helfand}, D.~J. 1995, \apj, 450, 559

\bibitem[{{Belsole} {et~al.}(2007){Belsole}, {Worrall}, {Hardcastle}, \&
  {Croston}}]{Belsole07}
{Belsole}, E., {Worrall}, D.~M., {Hardcastle}, M.~J., \& {Croston}, J.~H. 2007,
  \mnras, 381, 1109

\bibitem[{{Bertin} \& {Arnouts}(1996)}]{BA96}
{Bertin}, E., \& {Arnouts}, S. 1996, \aaps, 117, 393

\bibitem[{{Best}(2004)}]{Best04}
{Best}, P.~N. 2004, \mnras, 351, 70

\bibitem[{{Best} {et~al.}(2005){Best}, {Kauffmann}, {Heckman}, {Brinchmann},
  {Charlot}, {Ivezi{\'c}}, \& {White}}]{Best05}
{Best}, P.~N., {Kauffmann}, G., {Heckman}, T.~M., {et~al.} 2005, \mnras, 362,
  25

\bibitem[{{Best} {et~al.}(2007){Best}, {von der Linden}, {Kauffmann},
  {Heckman}, \& {Kaiser}}]{Best07}
{Best}, P.~N., {von der Linden}, A., {Kauffmann}, G., {Heckman}, T.~M., \&
  {Kaiser}, C.~R. 2007, \mnras, 379, 894

\bibitem[{{Blanton} {et~al.}(2010){Blanton}, {Clarke}, {Sarazin}, {Randall}, \&
  {McNamara}}]{Blanton10}
{Blanton}, E.~L., {Clarke}, T.~E., {Sarazin}, C.~L., {Randall}, S.~W., \&
  {McNamara}, B.~R. 2010, Proceedings of the National Academy of Science, 107,
  7174

\bibitem[{{Bufanda} {et~al.}(2017){Bufanda}, {Hollowood}, {Jeltema}, {Rykoff},
  {Rozo}, {Martini}, {Abbott}, {Abdalla}, {Allam}, {Banerji},
  {Benoit-L{\'e}vy}, {Bertin}, {Brooks}, {Carnero Rosell}, {Carrasco Kind},
  {Carretero}, {Cunha}, {da Costa}, {Desai}, {Diehl}, {Dietrich}, {Evrard},
  {Fausti Neto}, {Flaugher}, {Frieman}, {Gerdes}, {Goldstein}, {Gruen},
  {Gruendl}, {Gutierrez}, {Honscheid}, {James}, {Kuehn}, {Kuropatkin}, {Lima},
  {Maia}, {Marshall}, {Melchior}, {Miquel}, {Mohr}, {Ogando}, {Plazas},
  {Romer}, {Rooney}, {Sanchez}, {Santiago}, {Scarpine}, {Sevilla-Noarbe},
  {Smith}, {Soares-Santos}, {Sobreira}, {Suchyta}, {Tarle}, {Thomas}, {Tucker},
  {Walker}, \& {DES Collaboration}}]{Bufanda17}
{Bufanda}, E., {Hollowood}, D., {Jeltema}, T.~E., {et~al.} 2017, \mnras, 465,
  2531

\bibitem[{{Castignani} {et~al.}(2014){Castignani}, {Chiaberge}, {Celotti},
  {Norman}, \& {De Zotti}}]{Castignani14}
{Castignani}, G., {Chiaberge}, M., {Celotti}, A., {Norman}, C., \& {De Zotti},
  G. 2014, \apj, 792, 114

\bibitem[{{Cavaliere} \& {Fusco-Femiano}(1978)}]{Cavaliere78}
{Cavaliere}, A., \& {Fusco-Femiano}, R. 1978, Astronomy and Astrophysics, 70,
  677

\bibitem[{{Chabrier}(2003)}]{Chab03}
{Chabrier}, G. 2003, \pasp, 115, 763

\bibitem[{{Chambers} {et~al.}(2016){Chambers}, {Magnier}, {Metcalfe},
  {Flewelling}, {Huber}, {Waters}, {Denneau}, {Draper}, {Farrow}, {Finkbeiner},
  {Holmberg}, {Koppenhoefer}, {Price}, {Saglia}, {Schlafly}, {Smartt},
  {Sweeney}, {Wainscoat}, {Burgett}, {Grav}, {Heasley}, {Hodapp}, {Jedicke},
  {Kaiser}, {Kudritzki}, {Luppino}, {Lupton}, {Monet}, {Morgan}, {Onaka},
  {Stubbs}, {Tonry}, {Banados}, {Bell}, {Bender}, {Bernard}, {Botticella},
  {Casertano}, {Chastel}, {Chen}, {Chen}, {Cole}, {Deacon}, {Frenk},
  {Fitzsimmons}, {Gezari}, {Goessl}, {Goggia}, {Goldman}, {Grebel}, {Hambly},
  {Hasinger}, {Heavens}, {Heckman}, {Henderson}, {Henning}, {Holman}, {Hopp},
  {Ip}, {Isani}, {Keyes}, {Koekemoer}, {Kotak}, {Long}, {Lucey}, {Liu},
  {Martin}, {McLean}, {Morganson}, {Murphy}, {Nieto-Santisteban}, {Norberg},
  {Peacock}, {Pier}, {Postman}, {Primak}, {Rae}, {Rest}, {Riess}, {Riffeser},
  {Rix}, {Roser}, {Schilbach}, {Schultz}, {Scolnic}, {Szalay}, {Seitz},
  {Shiao}, {Small}, {Smith}, {Soderblom}, {Taylor}, {Thakar}, {Thiel},
  {Thilker}, {Urata}, {Valenti}, {Walter}, {Watters}, {Werner}, {White},
  {Wood-Vasey}, \& {Wyse}}]{PS16}
{Chambers}, K.~C., {Magnier}, E.~A., {Metcalfe}, N., {et~al.} 2016, ArXiv
  e-prints, arXiv:1612.05560

\bibitem[{{Chiaberge} {et~al.}(2009){Chiaberge}, {Tremblay}, {Capetti},
  {Macchetto}, {Tozzi}, \& {Sparks}}]{Chiaberge09}
{Chiaberge}, M., {Tremblay}, G., {Capetti}, A., {et~al.} 2009, \apj, 696, 1103

\bibitem[{{Ching} {et~al.}(2017){Ching}, {Croom}, {Sadler}, {Robotham},
  {Brough}, {Baldry}, {Bland -Hawthorn}, {Colless}, {Driver}, {Holwerda},
  {Hopkins}, {Jarvis}, {Johnston}, {Kelvin}, {Liske}, {Loveday}, {Norberg},
  {Pracy}, {Steele}, {Thomas}, \& {Wang}}]{Ching17}
{Ching}, J.~H.~Y., {Croom}, S.~M., {Sadler}, E.~M., {et~al.} 2017, \mnras, 469,
  4584

\bibitem[{{Condon}(1992)}]{Condon92}
{Condon}, J.~J. 1992, \araa, 30, 575

\bibitem[{{Connor} {et~al.}(2019){Connor}, {Kelson}, {Donahue}, \&
  {Moustakas}}]{Connor19}
{Connor}, T., {Kelson}, D.~D., {Donahue}, M., \& {Moustakas}, J. 2019, \apj,
  875, 16

\bibitem[{{Conroy} \& {Gunn}(2010)}]{c10}
{Conroy}, C., \& {Gunn}, J.~E. 2010, \apj, 712, 833

\bibitem[{{Conroy} {et~al.}(2009){Conroy}, {Gunn}, \& {White}}]{c09}
{Conroy}, C., {Gunn}, J.~E., \& {White}, M. 2009, \apj, 699, 486

\bibitem[{{Croston} {et~al.}(2018){Croston}, {Ineson}, \&
  {Hardcastle}}]{Croston18}
{Croston}, J.~H., {Ineson}, J., \& {Hardcastle}, M.~J. 2018, Monthly Notices of
  the Royal Astronomical Society, 476, 1614

\bibitem[{{Croston} {et~al.}(2008){Croston}, {Pratt}, {B{\"o}hringer},
  {Arnaud}, {Pointecouteau}, {Ponman}, {Sanderson}, {Temple}, {Bower}, \&
  {Donahue}}]{Croston08}
{Croston}, J.~H., {Pratt}, G.~W., {B{\"o}hringer}, H., {et~al.} 2008, \aap,
  487, 431

\bibitem[{{Croston} {et~al.}(2019){Croston}, {Hardcastle}, {Mingo}, {Best},
  {Sabater}, {Shimwell}, {Williams}, {Duncan}, {R{\"o}ttgering}, {Brienza},
  {G{\"u}rkan}, {Ineson}, {Miley}, {Morabito}, {O'Sullivan}, \&
  {Prandoni}}]{Croston19}
{Croston}, J.~H., {Hardcastle}, M.~J., {Mingo}, B., {et~al.} 2019, \aap, 622,
  A10

\bibitem[{{Di Matteo} {et~al.}(2005){Di Matteo}, {Springel}, \&
  {Hernquist}}]{DiMatteo05}
{Di Matteo}, T., {Springel}, V., \& {Hernquist}, L. 2005, \nat, 433, 604

\bibitem[{{Fabian}(2012)}]{Fabian12}
{Fabian}, A.~C. 2012, \araa, 50, 455

\bibitem[{{Fabian} {et~al.}(2005){Fabian}, {Sanders}, {Taylor}, \&
  {Allen}}]{Fabian05}
{Fabian}, A.~C., {Sanders}, J.~S., {Taylor}, G.~B., \& {Allen}, S.~W. 2005,
  Monthly Notices of the Royal Astronomical Society, 360, L20

\bibitem[{{Fabian} {et~al.}(2000){Fabian}, {Sanders}, {Ettori}, {Taylor},
  {Allen}, {Crawford}, {Iwasawa}, {Johnstone}, \& {Ogle}}]{Fabian00}
{Fabian}, A.~C., {Sanders}, J.~S., {Ettori}, S., {et~al.} 2000, \mnras, 318,
  L65

\bibitem[{{Falder} {et~al.}(2010){Falder}, {Stevens}, {Jarvis}, {Hardcastle},
  {Lacy}, {McLure}, {Hatziminaoglou}, {Page}, \& {Richards}}]{Falder10}
{Falder}, J.~T., {Stevens}, J.~A., {Jarvis}, M.~J., {et~al.} 2010, \mnras, 405,
  347

\bibitem[{{Falle}(1991)}]{Falle91}
{Falle}, S.~A.~E.~G. 1991, \mnras, 250, 581

\bibitem[{{Fanaroff} \& {Riley}(1974)}]{FR74}
{Fanaroff}, B.~L., \& {Riley}, J.~M. 1974, \mnras, 167, 31P

\bibitem[{{Forman} {et~al.}(2007){Forman}, {Jones}, {Churazov}, {Markevitch},
  {Nulsen}, {Vikhlinin}, {Begelman}, {B{\"o}hringer}, {Eilek}, {Heinz},
  {Kraft}, {Owen}, \& {Pahre}}]{Forman07}
{Forman}, W., {Jones}, C., {Churazov}, E., {et~al.} 2007, The Astrophysical
  Journal, 665, 1057

\bibitem[{{Galametz} {et~al.}(2009){Galametz}, {Stern}, {Eisenhardt},
  {Brodwin}, {Brown}, {Dey}, {Gonzalez}, {Jannuzi}, {Moustakas}, \&
  {Stanford}}]{Galametz09}
{Galametz}, A., {Stern}, D., {Eisenhardt}, P. R.~M., {et~al.} 2009, \apj, 694,
  1309

\bibitem[{{Garon} {et~al.}(2019){Garon}, {Rudnick}, {Wong}, {Jones}, {Kim},
  {Andernach}, {Shabala}, {Kapi{\'n}ska}, {Norris}, {de Gasperin}, {Tate}, \&
  {Tang}}]{Garon19}
{Garon}, A.~F., {Rudnick}, L., {Wong}, O.~I., {et~al.} 2019, \aj, 157, 126

\bibitem[{{Gaspari} {et~al.}(2011){Gaspari}, {Melioli}, {Brighenti}, \&
  {D'Ercole}}]{Gaspari11}
{Gaspari}, M., {Melioli}, C., {Brighenti}, F., \& {D'Ercole}, A. 2011, Monthly
  Notices of the Royal Astronomical Society, 411, 349

\bibitem[{{Gendre} {et~al.}(2013){Gendre}, {Best}, {Wall}, \& {Ker}}]{Gendre13}
{Gendre}, M.~A., {Best}, P.~N., {Wall}, J.~V., \& {Ker}, L.~M. 2013, \mnras,
  430, 3086

\bibitem[{{Ginsburg} {et~al.}(2019){Ginsburg}, {Sip{\H{o}}cz}, {Brasseur},
  {Cowperthwaite}, {Craig}, {Deil}, {Guillochon}, {Guzman}, {Liedtke}, {Lian
  Lim}, {Lockhart}, {Mommert}, {Morris}, {Norman}, {Parikh}, {Persson},
  {Robitaille}, {Segovia}, {Singer}, {Tollerud}, {de Val-Borro}, {Valtchanov},
  {Woillez}, {Astroquery Collaboration}, \& {a subset of the astropy
  Collaboration}}]{Ginsburg19}
{Ginsburg}, A., {Sip{\H{o}}cz}, B.~M., {Brasseur}, C.~E., {et~al.} 2019, \aj,
  157, 98

\bibitem[{{Gonzalez} {et~al.}(2019){Gonzalez}, {Gettings}, {Brodwin},
  {Eisenhardt}, {Stanford}, {Wylezalek}, {Decker}, {Marrone}, {Moravec},
  {O'Donnell}, {Stalder}, {Stern}, {Abdulla}, {Brown}, {Carlstrom}, {Chambers},
  {Hayden}, {Lin}, {Magnier}, {Masci}, {Mantz}, {McDonald}, {Mo}, {Perlmutter},
  {Wright}, \& {Zeimann}}]{Gonzalez19}
{Gonzalez}, A.~H., {Gettings}, D.~P., {Brodwin}, M., {et~al.} 2019, The
  Astrophysical Journal Supplement Series, 240, 33

\bibitem[{{Gralla} {et~al.}(2011){Gralla}, {Gladders}, {Yee}, \&
  {Barrientos}}]{Gralla11}
{Gralla}, M.~B., {Gladders}, M.~D., {Yee}, H.~K.~C., \& {Barrientos}, L.~F.
  2011, \apj, 734, 103

\bibitem[{{Granato} {et~al.}(2004){Granato}, {De Zotti}, {Silva}, {Bressan}, \&
  {Danese}}]{Granato04}
{Granato}, G.~L., {De Zotti}, G., {Silva}, L., {Bressan}, A., \& {Danese}, L.
  2004, \apj, 600, 580

\bibitem[{{Gupta} {et~al.}(2019){Gupta}, {Pannella}, {Mohr}, {Klein}, {Rykoff},
  {Annis}, {Avila}, {Bianchini}, {Brooks}, {Buckley-Geer}, {Bulbul}, {Carnero
  Rosell}, {Carrasco Kind}, {Carretero}, {Chiu}, {Costanzi}, {da Costa}, {De
  Vicente}, {Desai}, {Dietrich}, {Doel}, {Everett}, {Evrard},
  {Garc{\'\i}a-Bellido}, {Gaztanaga}, {Gruendl}, {Gschwend}, {Gutierrez},
  {Hollowood}, {Honscheid}, {James}, {Jeltema}, {Kuehn}, {Lidman}, {Lima},
  {Maia}, {Marshall}, {McDonald}, {Menanteau}, {Miquel}, {Ogand o}, {Palmese},
  {Paz-Chinch{\'o}n}, {Plazas}, {Reichardt}, {Sanchez}, {Santiago}, {Saro},
  {Scarpine}, {Schindler}, {Schubnell}, {Serrano}, {Sevilla-Noarbe}, {Shao},
  {Smith}, {Stott}, {Strazzullo}, {Suchyta}, {Swanson}, {Vikram}, \&
  {Zenteno}}]{Gupta19}
{Gupta}, N., {Pannella}, M., {Mohr}, J.~J., {et~al.} 2019, arXiv e-prints,
  arXiv:1906.11388

\bibitem[{{Hlavacek-Larrondo} {et~al.}(2015){Hlavacek-Larrondo}, {McDonald},
  {Benson}, {Forman}, {Allen}, {Bleem}, {Ashby}, {Bocquet}, {Brodwin},
  {Dietrich}, {Jones}, {Liu}, {Reichardt}, {Saliwanchik}, {Saro}, {Schrabback},
  {Song}, {Stalder}, {Vikhlinin}, \& {Zenteno}}]{HL15}
{Hlavacek-Larrondo}, J., {McDonald}, M., {Benson}, B.~A., {et~al.} 2015, \apj,
  805, 35

\bibitem[{{Hopkins} \& {Elvis}(2010)}]{Hopkins10}
{Hopkins}, P.~F., \& {Elvis}, M. 2010, \mnras, 401, 7

\bibitem[{{Ineson} {et~al.}(2013){Ineson}, {Croston}, {Hardcastle}, {Kraft},
  {Evans}, \& {Jarvis}}]{Ineson13}
{Ineson}, J., {Croston}, J.~H., {Hardcastle}, M.~J., {et~al.} 2013, \apj, 770,
  136

\bibitem[{{Ineson} {et~al.}(2015){Ineson}, {Croston}, {Hardcastle}, {Kraft},
  {Evans}, \& {Jarvis}}]{Ineson15}
---. 2015, \mnras, 453, 2682

\bibitem[{{Janssen} {et~al.}(2012){Janssen}, {R{\"o}ttgering}, {Best}, \&
  {Brinchmann}}]{Janssen12}
{Janssen}, R.~M.~J., {R{\"o}ttgering}, H.~J.~A., {Best}, P.~N., \&
  {Brinchmann}, J. 2012, \aap, 541, A62

\bibitem[{{Johnstone} {et~al.}(2002){Johnstone}, {Allen}, {Fabian}, \& {Sand
  ers}}]{Johnstone02}
{Johnstone}, R.~M., {Allen}, S.~W., {Fabian}, A.~C., \& {Sand ers}, J.~S. 2002,
  Monthly Notices of the Royal Astronomical Society, 336, 299

\bibitem[{{Kaiser} \& {Alexander}(1997)}]{Kaiser97}
{Kaiser}, C.~R., \& {Alexander}, P. 1997, \mnras, 286, 215

\bibitem[{{Kaiser} \& {Best}(2007)}]{Kaiser07}
{Kaiser}, C.~R., \& {Best}, P.~N. 2007, Monthly Notices of the Royal
  Astronomical Society, 381, 1548

\bibitem[{{Kauffmann} \& {Haehnelt}(2000)}]{Kauffmann00}
{Kauffmann}, G., \& {Haehnelt}, M. 2000, \mnras, 311, 576

\bibitem[{{Kellermann} \& {Owen}(1988)}]{Kellermann88}
{Kellermann}, K.~I., \& {Owen}, F.~N. 1988, {Radio galaxies and quasars}
  (Springer-Verlag), 563--602

\bibitem[{{Ledlow} \& {Owen}(1996)}]{LO96}
{Ledlow}, M.~J., \& {Owen}, F.~N. 1996, \aj, 112, 9

\bibitem[{{Li} {et~al.}(2017){Li}, {Ruszkowski}, \& {Bryan}}]{Li17}
{Li}, Y., {Ruszkowski}, M., \& {Bryan}, G.~L. 2017, The Astrophysical Journal,
  847, 106

\bibitem[{{Lin} \& {Mohr}(2007)}]{L&M07}
{Lin}, Y.-T., \& {Mohr}, J.~J. 2007, \apjs, 170, 71

\bibitem[{{Makovoz} \& {Khan}(2005)}]{MK05}
{Makovoz}, D., \& {Khan}, I. 2005, in Astronomical Society of the Pacific
  Conference Series, Vol. 347, Astronomical Data Analysis Software and Systems
  XIV, ed. P.~{Shopbell}, M.~{Britton}, \& R.~{Ebert}, 81

\bibitem[{{Mancone} \& {Gonzalez}(2012)}]{EzGal12}
{Mancone}, C.~L., \& {Gonzalez}, A.~H. 2012, \pasp, 124, 606

\bibitem[{{Martini} {et~al.}(2013){Martini}, {Miller}, {Brodwin}, {Stanford},
  {Gonzalez}, {Bautz}, {Hickox}, {Stern}, {Eisenhardt}, {Galametz}, {Norman},
  {Jannuzi}, {Dey}, {Murray}, {Jones}, \& {Brown}}]{Martini13}
{Martini}, P., {Miller}, E.~D., {Brodwin}, M., {et~al.} 2013, \apj, 768, 1

\bibitem[{{McDonald} {et~al.}(2018){McDonald}, {Gaspari}, {McNamara}, \&
  {Tremblay}}]{McDonald18}
{McDonald}, M., {Gaspari}, M., {McNamara}, B.~R., \& {Tremblay}, G.~R. 2018,
  The Astrophysical Journal, 858, 45

\bibitem[{{McDonald} {et~al.}(2013){McDonald}, {Benson}, {Vikhlinin},
  {Stalder}, {Bleem}, {de Haan}, {Lin}, {Aird}, {Ashby}, {Bautz}, {Bayliss},
  {Bocquet}, {Brodwin}, {Carlstrom}, {Chang}, {Cho}, {Clocchiatti}, {Crawford},
  {Crites}, {Desai}, {Dobbs}, {Dudley}, {Foley}, {Forman}, {George},
  {Gettings}, {Gladders}, {Gonzalez}, {Halverson}, {High}, {Holder},
  {Holzapfel}, {Hoover}, {Hrubes}, {Jones}, {Joy}, {Keisler}, {Knox}, {Lee},
  {Leitch}, {Liu}, {Lueker}, {Luong-Van}, {Mantz}, {Marrone}, {McMahon},
  {Mehl}, {Meyer}, {Miller}, {Mocanu}, {Mohr}, {Montroy}, {Murray},
  {Nurgaliev}, {Padin}, {Plagge}, {Pryke}, {Reichardt}, {Rest}, {Ruel}, {Ruhl},
  {Saliwanchik}, {Saro}, {Sayre}, {Schaffer}, {Shirokoff}, {Song}, {{\v
  S}uhada}, {Spieler}, {Stanford}, {Staniszewski}, {Stark}, {Story}, {van
  Engelen}, {Vanderlinde}, {Vieira}, {Williamson}, {Zahn}, \&
  {Zenteno}}]{McDonald13}
{McDonald}, M., {Benson}, B.~A., {Vikhlinin}, A., {et~al.} 2013, \apj, 774, 23

\bibitem[{{McMullin} {et~al.}(2007){McMullin}, {Waters}, {Schiebel}, {Young},
  \& {Golap}}]{casa}
{McMullin}, J.~P., {Waters}, B., {Schiebel}, D., {Young}, W., \& {Golap}, K.
  2007, in Astronomical Society of the Pacific Conference Series, Vol. 376,
  Astronomical Data Analysis Software and Systems XVI, ed. R.~A. {Shaw},
  F.~{Hill}, \& D.~J. {Bell}, 127

\bibitem[{{McNamara} \& {Nulsen}(2007)}]{McNamara07}
{McNamara}, B.~R., \& {Nulsen}, P.~E.~J. 2007, Annual Review of Astronomy and
  Astrophysics, 45, 117

\bibitem[{{McNamara} {et~al.}(2000){McNamara}, {Wise}, {Nulsen}, {David},
  {Sarazin}, {Bautz}, {Markevitch}, {Vikhlinin}, {Forman}, {Jones}, \&
  {Harris}}]{McNamara00}
{McNamara}, B.~R., {Wise}, M., {Nulsen}, P.~E.~J., {et~al.} 2000, The
  Astrophysical Journal, 534, L135

\bibitem[{{Miley} \& {De Breuck}(2008)}]{Miley08}
{Miley}, G., \& {De Breuck}, C. 2008, \aapr, 15, 67

\bibitem[{{Mingo} {et~al.}(2019){Mingo}, {Croston}, {Hardcastle}, {Best},
  {Duncan}, {Morganti}, {Rottgering}, {Sabater}, {Shimwell}, {Williams},
  {Brienza}, {Gurkan}, {Mahatma}, {Morabito}, {Prandoni}, {Bondi}, {Ineson}, \&
  {Mooney}}]{Mingo19}
{Mingo}, B., {Croston}, J.~H., {Hardcastle}, M.~J., {et~al.} 2019, \mnras, 488,
  2701

\bibitem[{{Minkowski}(1960)}]{Minkowski60}
{Minkowski}, R. 1960, \apj, 132, 908

\bibitem[{{Mo} {et~al.}(2018){Mo}, {Gonzalez}, {Stern}, {Brodwin}, {Decker},
  {Eisenhardt}, {Moravec}, {Stanford}, \& {Wylezalek}}]{Mo18}
{Mo}, W., {Gonzalez}, A., {Stern}, D., {et~al.} 2018, \apj, 869, 131

\bibitem[{{Moravec} {et~al.}(2019){Moravec}, {Gonzalez}, {Stern}, {Brodwin},
  {Clarke}, {Decker}, {Eisenhardt}, {Mo}, {O{\textquoteright}Donnell}, {Pope},
  {Stanford}, \& {Wylezalek}}]{EM19}
{Moravec}, E., {Gonzalez}, A.~H., {Stern}, D., {et~al.} 2019, \apj, 871, 186

\bibitem[{{Noirot} {et~al.}(2016){Noirot}, {Vernet}, {De Breuck}, {Wylezalek},
  {Galametz}, {Stern}, {Mei}, {Brodwin}, {Cooke}, {Gonzalez}, {Hatch},
  {Rettura}, \& {Stanford}}]{Noirot16}
{Noirot}, G., {Vernet}, J., {De Breuck}, C., {et~al.} 2016, \apj, 830, 90

\bibitem[{{Noirot} {et~al.}(2018){Noirot}, {Stern}, {Mei}, {Wylezalek},
  {Cooke}, {De Breuck}, {Galametz}, {Hatch}, {Vernet}, {Brodwin}, {Eisenhardt},
  {Gonzalez}, {Jarvis}, {Rettura}, {Seymour}, \& {Stanford}}]{Noirot18}
{Noirot}, G., {Stern}, D., {Mei}, S., {et~al.} 2018, \apj, 859, 38

\bibitem[{{O'Dea} {et~al.}(2008){O'Dea}, {Baum}, {Privon}, {Noel-Storr},
  {Quillen}, {Zufelt}, {Park}, {Edge}, {Russell}, {Fabian}, {Donahue},
  {Sarazin}, {McNamara}, {Bregman}, \& {Egami}}]{ODea08}
{O'Dea}, C.~P., {Baum}, S.~A., {Privon}, G., {et~al.} 2008, The Astrophysical
  Journal, 681, 1035

\bibitem[{{Paterno-Mahler} {et~al.}(2017){Paterno-Mahler}, {Blanton},
  {Brodwin}, {Ashby}, {Golden-Marx}, {Decker}, {Wing}, \& {Anand}}]{PM17}
{Paterno-Mahler}, R., {Blanton}, E.~L., {Brodwin}, M., {et~al.} 2017, \apj,
  844, 78

\bibitem[{{Peterson}(1997)}]{PetersonBook}
{Peterson}, B.~M. 1997, {An Introduction to Active Galactic Nuclei} (Cambridge
  University Press)

\bibitem[{{Planck Collaboration} {et~al.}(2016){Planck Collaboration}, {Ade},
  {Aghanim}, {Arnaud}, {Ashdown}, {Aumont}, {Baccigalupi}, {Banday},
  {Barreiro}, {Bartlett}, {Bartolo}, {Battaner}, {Battye}, {Benabed},
  {Beno{\^\i}t}, {Benoit-L{\'e}vy}, {Bernard}, {Bersanelli}, {Bielewicz},
  {Bock}, {Bonaldi}, {Bonavera}, {Bond}, {Borrill}, {Bouchet}, {Boulanger},
  {Bucher}, {Burigana}, {Butler}, {Calabrese}, {Cardoso}, {Catalano},
  {Challinor}, {Chamballu}, {Chary}, {Chiang}, {Chluba}, {Christensen},
  {Church}, {Clements}, {Colombi}, {Colombo}, {Combet}, {Coulais}, {Crill},
  {Curto}, {Cuttaia}, {Danese}, {Davies}, {Davis}, {de Bernardis}, {de Rosa},
  {de Zotti}, {Delabrouille}, {D{\'e}sert}, {Di Valentino}, {Dickinson},
  {Diego}, {Dolag}, {Dole}, {Donzelli}, {Dor{\'e}}, {Douspis}, {Ducout},
  {Dunkley}, {Dupac}, {Efstathiou}, {Elsner}, {En{\ss}lin}, {Eriksen},
  {Farhang}, {Fergusson}, {Finelli}, {Forni}, {Frailis}, {Fraisse},
  {Franceschi}, {Frejsel}, {Galeotta}, {Galli}, {Ganga}, {Gauthier}, {Gerbino},
  {Ghosh}, {Giard}, {Giraud-H{\'e}raud}, {Giusarma}, {Gjerl{\o}w},
  {Gonz{\'a}lez-Nuevo}, {G{\'o}rski}, {Gratton}, {Gregorio}, {Gruppuso},
  {Gudmundsson}, {Hamann}, {Hansen}, {Hanson}, {Harrison}, {Helou}, {Henrot-
  Versill{\'e}}, {Hern{\'a}ndez-Monteagudo}, {Herranz}, {Hildebrandt}, {Hivon},
  {Hobson}, {Holmes}, {Hornstrup}, {Hovest}, {Huang}, {Huffenberger}, {Hurier},
  {Jaffe}, {Jaffe}, {Jones}, {Juvela}, {Keih{\"a}nen}, {Keskitalo}, {Kisner},
  {Kneissl}, {Knoche}, {Knox}, {Kunz}, {Kurki-Suonio}, {Lagache},
  {L{\"a}hteenm{\"a}ki}, {Lamarre}, {Lasenby}, {Lattanzi}, {Lawrence}, {Leahy},
  {Leonardi}, {Lesgourgues}, {Levrier}, {Lewis}, {Liguori}, {Lilje},
  {Linden-V{\o}rnle}, {L{\'o}pez-Caniego}, {Lubin}, {Mac{\'\i}as-P{\'e}rez},
  {Maggio}, {Maino}, {Mandolesi}, {Mangilli}, {Marchini}, {Maris}, {Martin},
  {Martinelli}, {Mart{\'\i}nez-Gonz{\'a}lez}, {Masi}, {Matarrese}, {McGehee},
  {Meinhold}, {Melchiorri}, {Melin}, {Mendes}, {Mennella}, {Migliaccio},
  {Millea}, {Mitra}, {Miville-Desch{\^e}nes}, {Moneti}, {Montier}, {Morgante},
  {Mortlock}, {Moss}, {Munshi}, {Murphy}, {Naselsky}, {Nati}, {Natoli},
  {Netterfield}, {N{\o}rgaard-Nielsen}, {Noviello}, {Novikov}, {Novikov},
  {Oxborrow}, {Paci}, {Pagano}, {Pajot}, {Paladini}, {Paoletti}, {Partridge},
  {Pasian}, {Patanchon}, {Pearson}, {Perdereau}, {Perotto}, {Perrotta},
  {Pettorino}, {Piacentini}, {Piat}, {Pierpaoli}, {Pietrobon}, {Plaszczynski},
  {Pointecouteau}, {Polenta}, {Popa}, {Pratt}, {Pr{\'e}zeau}, {Prunet},
  {Puget}, {Rachen}, {Reach}, {Rebolo}, {Reinecke}, {Remazeilles}, {Renault},
  {Renzi}, {Ristorcelli}, {Rocha}, {Rosset}, {Rossetti}, {Roudier},
  {Rouill{\'e} d'Orfeuil}, {Rowan-Robinson}, {Rubi{\~n}o-Mart{\'\i}n},
  {Rusholme}, {Said}, {Salvatelli}, {Salvati}, {Sandri}, {Santos},
  {Savelainen}, {Savini}, {Scott}, {Seiffert}, {Serra}, {Shellard}, {Spencer},
  {Spinelli}, {Stolyarov}, {Stompor}, {Sudiwala}, {Sunyaev}, {Sutton},
  {Suur-Uski}, {Sygnet}, {Tauber}, {Terenzi}, {Toffolatti}, {Tomasi},
  {Tristram}, {Trombetti}, {Tucci}, {Tuovinen}, {T{\"u}rler}, {Umana},
  {Valenziano}, {Valiviita}, {Van Tent}, {Vielva}, {Villa}, {Wade}, {Wandelt},
  {Wehus}, {White}, {White}, {Wilkinson}, {Yvon}, {Zacchei}, \&
  {Zonca}}]{Planck15}
{Planck Collaboration}, {Ade}, P.~A.~R., {Aghanim}, N., {et~al.} 2016, \aap,
  594, A13

\bibitem[{{Prasad} {et~al.}(2015){Prasad}, {Sharma}, \& {Babul}}]{Prasad15}
{Prasad}, D., {Sharma}, P., \& {Babul}, A. 2015, The Astrophysical Journal,
  811, 108

\bibitem[{{Quintana}(1979)}]{Quintana79}
{Quintana}, H. 1979, \aj, 84, 15

\bibitem[{{Rigby} {et~al.}(2014){Rigby}, {Hatch}, {R{\"o}ttgering},
  {Sibthorpe}, {Chiang}, {Overzier}, {Herbonnet}, {Borgani}, {Clements},
  {Dannerbauer}, {De Breuck}, {De Lucia}, {Kurk}, {Maschietto}, {Miley},
  {Saro}, {Seymour}, \& {Venemans}}]{Rigby14}
{Rigby}, E.~E., {Hatch}, N.~A., {R{\"o}ttgering}, H.~J.~A., {et~al.} 2014,
  \mnras, 437, 1882

\bibitem[{{Roberts} {et~al.}(2015){Roberts}, {Parker}, {Joshi}, \&
  {Evans}}]{Roberts15}
{Roberts}, I.~D., {Parker}, L.~C., {Joshi}, G.~D., \& {Evans}, F.~A. 2015,
  \mnras, 448, L1

\bibitem[{{Robitaille} \& {Bressert}(2012)}]{aplpy}
{Robitaille}, T., \& {Bressert}, E. 2012, {APLpy: Astronomical Plotting Library
  in Python}, Astrophysics Source Code Library, ascl:1208.017

\bibitem[{{Sabater} {et~al.}(2019){Sabater}, {Best}, {Hardcastle}, {Shimwell},
  {Tasse}, {Williams}, {Br{\"u}ggen}, {Cochrane}, {Croston}, {de Gasperin},
  {Duncan}, {G{\"u}rkan}, {Mechev}, {Morabito}, {Prandoni}, {R{\"o}ttgering},
  {Smith}, {Harwood}, {Mingo}, {Mooney}, \& {Saxena}}]{Sabater19}
{Sabater}, J., {Best}, P.~N., {Hardcastle}, M.~J., {et~al.} 2019, \aap, 622,
  A17

\bibitem[{{Schuster} {et~al.}(2006){Schuster}, {Marengo}, \&
  {Patten}}]{Schuster06}
{Schuster}, M.~T., {Marengo}, M., \& {Patten}, B.~M. 2006, in \procspie, Vol.
  6270, Society of Photo-Optical Instrumentation Engineers (SPIE) Conference
  Series, 627020

\bibitem[{{Seymour} {et~al.}(2007){Seymour}, {Stern}, {De Breuck}, {Vernet},
  {Rettura}, {Dickinson}, {Dey}, {Eisenhardt}, {Fosbury}, {Lacy}, {McCarthy},
  {Miley}, {Rocca-Volmerange}, {R{\"o}ttgering}, {Stanford}, {Teplitz}, {van
  Breugel}, \& {Zirm}}]{Seymour07}
{Seymour}, N., {Stern}, D., {De Breuck}, C., {et~al.} 2007, \apjs, 171, 353

\bibitem[{{Shimwell} {et~al.}(2019){Shimwell}, {Tasse}, {Hardcastle}, {Mechev},
  {Williams}, {Best}, {R{\"o}ttgering}, {Callingham}, {Dijkema}, {de Gasperin},
  {Hoang}, {Hugo}, {Mirmont}, {Oonk}, {Prandoni}, {Rafferty}, {Sabater},
  {Smirnov}, {van Weeren}, {White}, {Atemkeng}, {Bester}, {Bonnassieux},
  {Br{\"u}ggen}, {Brunetti}, {Chy{\.z}y}, {Cochrane}, {Conway}, {Croston},
  {Danezi}, {Duncan}, {Haverkorn}, {Heald}, {Iacobelli}, {Intema}, {Jackson},
  {Jamrozy}, {Jarvis}, {Lakhoo}, {Mevius}, {Miley}, {Morabito}, {Morganti},
  {Nisbet}, {Orr{\'u}}, {Perkins}, {Pizzo}, {Schrijvers}, {Smith}, {Vermeulen},
  {Wise}, {Alegre}, {Bacon}, {van Bemmel}, {Beswick}, {Bonafede}, {Botteon},
  {Bourke}, {Brienza}, {Calistro Rivera}, {Cassano}, {Clarke}, {Conselice},
  {Dettmar}, {Drabent}, {Dumba}, {Emig}, {En{\ss}lin}, {Ferrari}, {Garrett},
  {G{\'e}nova-Santos}, {Goyal}, {G{\"u}rkan}, {Hale}, {Harwood}, {Heesen},
  {Hoeft}, {Horellou}, {Jackson}, {Kokotanekov}, {Kondapally},
  {Kunert-Bajraszewska}, {Mahatma}, {Mahony}, {Mandal}, {McKean}, {Merloni},
  {Mingo}, {Miskolczi}, {Mooney}, {Nikiel-Wroczy{\'n}ski}, {O'Sullivan},
  {Quinn}, {Reich}, {Roskowi{\'n}ski}, {Rowlinson}, {Savini}, {Saxena},
  {Schwarz}, {Shulevski}, {Sridhar}, {Stacey}, {Urquhart}, {van der Wiel},
  {Varenius}, {Webster}, \& {Wilber}}]{Shimwell19}
{Shimwell}, T.~W., {Tasse}, C., {Hardcastle}, M.~J., {et~al.} 2019, \aap, 622,
  A1

\bibitem[{{Silverstein} {et~al.}(2018){Silverstein}, {Anderson}, \&
  {Bregman}}]{Silverstein18}
{Silverstein}, E.~M., {Anderson}, M.~E., \& {Bregman}, J.~N. 2018, \aj, 155, 14

\bibitem[{{Stern} {et~al.}(2000){Stern}, {Djorgovski}, {Perley}, {de Carvalho},
  \& {Wall}}]{Stern00}
{Stern}, D., {Djorgovski}, S.~G., {Perley}, R.~A., {de Carvalho}, R.~R., \&
  {Wall}, J.~V. 2000, \aj, 119, 1526

\bibitem[{{Stern} {et~al.}(2012){Stern}, {Assef}, {Benford}, {Blain}, {Cutri},
  {Dey}, {Eisenhardt}, {Griffith}, {Jarrett}, {Lake}, {Masci}, {Petty},
  {Stanford}, {Tsai}, {Wright}, {Yan}, {Harrison}, \& {Madsen}}]{Stern12}
{Stern}, D., {Assef}, R.~J., {Benford}, D.~J., {et~al.} 2012, \apj, 753, 30

\bibitem[{{Tchekhovskoy} \& {Bromberg}(2016)}]{TO16}
{Tchekhovskoy}, A., \& {Bromberg}, O. 2016, \mnras, 461, L46

\bibitem[{{Tiwari}(2016)}]{Tiwari16}
{Tiwari}, P. 2016, ArXiv e-prints, arXiv:1609.01308

\bibitem[{{Vikhlinin} {et~al.}(2006){Vikhlinin}, {Kravtsov}, {Forman}, {Jones},
  {Markevitch}, {Murray}, \& {Van Speybroeck}}]{Vikhlinin06}
{Vikhlinin}, A., {Kravtsov}, A., {Forman}, W., {et~al.} 2006, The Astrophysical
  Journal, 640, 691

\bibitem[{{Wold} {et~al.}(2000){Wold}, {Lacy}, {Lilje}, \& {Serjeant}}]{Wold00}
{Wold}, M., {Lacy}, M., {Lilje}, P.~B., \& {Serjeant}, S. 2000, \mnras, 316,
  267

\bibitem[{{Wright} {et~al.}(2010){Wright}, {Eisenhardt}, {Mainzer}, {Ressler},
  {Cutri}, {Jarrett}, {Kirkpatrick}, {Padgett}, {McMillan}, {Skrutskie},
  {Stanford}, {Cohen}, {Walker}, {Mather}, {Leisawitz}, {Gautier}, {McLean},
  {Benford}, {Lonsdale}, {Blain}, {Mendez}, {Irace}, {Duval}, {Liu}, {Royer},
  {Heinrichsen}, {Howard}, {Shannon}, {Kendall}, {Walsh}, {Larsen}, {Cardon},
  {Schick}, {Schwalm}, {Abid}, {Fabinsky}, {Naes}, \& {Tsai}}]{Wright10}
{Wright}, E.~L., {Eisenhardt}, P. R.~M., {Mainzer}, A.~K., {et~al.} 2010, \aj,
  140, 1868

\bibitem[{{Wylezalek} {et~al.}(2013){Wylezalek}, {Galametz}, {Stern}, {Vernet},
  {De Breuck}, {Seymour}, {Brodwin}, {Eisenhardt}, {Gonzalez}, {Hatch},
  {Jarvis}, {Rettura}, {Stanford}, \& {Stevens}}]{Wylezalek13}
{Wylezalek}, D., {Galametz}, A., {Stern}, D., {et~al.} 2013, \apj, 769, 79

\bibitem[{{Wylezalek} {et~al.}(2014){Wylezalek}, {Vernet}, {De Breuck},
  {Stern}, {Brodwin}, {Galametz}, {Gonzalez}, {Jarvis}, {Hatch}, {Seymour}, \&
  {Stanford}}]{Wylezalek14}
{Wylezalek}, D., {Vernet}, J., {De Breuck}, C., {et~al.} 2014, \apj, 786, 17

\bibitem[{{Yang} {et~al.}(2019){Yang}, {Gaspari}, \& {Marlow}}]{Yang19}
{Yang}, H. Y.~K., {Gaspari}, M., \& {Marlow}, C. 2019, The Astrophysical
  Journal, 871, 6

\bibitem[{{Yuan} {et~al.}(2016){Yuan}, {Han}, \& {Wen}}]{Yuan16}
{Yuan}, Z.~S., {Han}, J.~L., \& {Wen}, Z.~L. 2016, \mnras, 460, 3669

\end{thebibliography}
\bibliographystyle{apj}

\end{document}